\documentclass{ptptex}

\usepackage{amsmath,amssymb}
\usepackage[dvipdfm]{graphicx}
\usepackage{subfigure}

\title{A new approach to investigate dineutron correlation \\ and its application to $^{10}$Be} 

\author{Fumiharu \textsc{Kobayashi} and Yoshiko \textsc{Kanada-En'yo}}

\inst{Department of Physics, Kyoto University, Kyoto 606-8502, Japan}

\abst{We propose a new framework by means of the dineutron condensate (DC) wave function
to describe the dineutron correlation, which is characterized by the spatially strong correlation of 
a spin-zero neutron-neutron pair, in neutron-rich nuclei with an active deformed core 
surrounded by valence neutrons. 
Using the DC wave function for a 2$\alpha$+2n system, which corresponds to a toy model for 
the $^{10}$Be system, we investigate the neutron-neutron correlation 
around the $2\alpha$ core and discuss the mechanism of the dineutron formation 
at the surface of finite nuclei.
To investigate dineutron correlations in realistic nuclear systems, 
we superpose the antisymmetrized molecular dynamics (AMD) wave functions and the DC wave functions.
Applying the AMD+DC method to $^{10}$Be, 
we show effects of the DC wave functions in the ground and excited $0^+$ states of $^{10}$Be 
and discuss the dineutron correlation in them.}

\begin{document}

\maketitle

\section{Introduction}

Recently, neutron-rich unstable nuclei are investigated intensively 
and various exotic phenomena have been discovered.
The dineutron correlation is one of the key issues attracting a great interest 
in physics of unstable nuclei.
The dineutron correlation means the strong spatial correlation 
between two neutrons in the spin-zero channel. 
Even though two neutrons do not form a bound state in free space, 
they attract each other rather strongly. 
It was suggested that the spatial correlations between two neutrons enhance
in a low-density infinite nuclear matter 
\cite{deblasio97,matsuo06,margueron07}, 
in the neutron-skin structure of medium-heavy nuclei \cite{matsuo05,pillet07} 
and also in the halo structure of light nuclei \cite{hansen87,bertsch91,zhukov93,
esbensen97,aoyama01,ikeda02,myo02,myo03,hagino05}. 
In such nuclear systems, the strongly correlated neutron-pair with a rather compact size 
can be regarded as a semi-bound or a virtually bound state called a dineutron.
The dineutron correlation is considered to be an important aspect 
to understand the structures of unstable nuclei.

In studies of the asymmetric nuclear matter,
it was shown that the strength of the two-neutron correlation changes 
as a function of the matter density 
\cite{bardo90,takatsuka93,deblasio97,dean03,matsuo06,
margueron07,isayev08}.
Matsuo suggested the continuous transitions 
from a weak BCS-like coupling to a strong boson-like coupling 
in accordance with a decrease of the matter density \cite{matsuo06}
(the BCS-BEC crossover, such as in a superfluid Fermi gas \cite{regal04}).
The bosonic behavior of dineutrons is often discussed 
in relation to the $\alpha$ condensation suggested 
in a low-density symmetric nuclear matter\cite{ropke98}.

The dineutron correlation in neutron-rich nuclei has been investigated well so far. 
In the medium-heavy nuclei, Matsuo {\it et al}. have investigated 
the neutron-neutron correlation in O, Ca, Ni isotopes \cite{matsuo05}. 
They used the HFB theory to deal with the correlation between two neutrons in those nuclei, 
and concluded that two neutrons in the neutron-skin region show the spatial correlation. 
Also in light-mass nuclei such as $^8$He, 
the dineutron correlation has been theoretically studied
\cite{enyo07,itagaki08,hagino08}.

For the neutron-halo nuclei, such as $^6$He and $^{11}$Li, 
where two valence neutrons are weakly bound to a core nucleus, 
the dineutron component was explicitly investigated 
by means of three-body models assuming a core and two valence neutrons 
\cite{hansen87,bertsch91,zhukov93,esbensen97,csoto93,baye94,
aoyama01,ikeda02,arai01,myo02,myo03,hagino05}.
In Ref.~\citen{hagino05}, 
Hagino {\it et al}. suggested that the pair of neutrons having
the strong angular correlation is distributed 
in the halo region of $^6$He and $^{11}$Li. 
In addition, they also have discussed the change in the correlation between two valence neutrons 
as a function of the distance from the core,
associating with the BCS-BEC crossover in finite nuclear systems \cite{hagino07}. 
Since most of the three-body models, however, have such an assumption 
as a spherical and inert core, 
they are not able to deal with systems having a deformed core and
their applications are limited to some specific nuclei.

In order to elucidate the properties and mechanisms 
of the dineutrons and their condensate states, 
it is significant to investigate dineutron behaviors in neutron-rich nuclei systematically. 
We, accordingly, have constructed a new model 
which can be applied to various nuclei having deformed structures
to draw the universal properties of dineutrons at the nuclear surface. 

In this model, the nuclear system is assumed to consist of 
a deformed core and surrounding dineutrons.  
The deformed core is described with antisymmetrized molecular dynamics (AMD)
\cite{enyo95,enyo01,enyo03},
and several pairs of neutrons around the core are written by the condensate wave function, 
in which dineutrons are moving in the $S$-orbit 
as $\alpha$s in the $\alpha$ condensate wave function \cite{tohsaki01}. 
We call this model ``the dineutron condensate (DC) wave function''.
The AMD is a useful model to describe structures of light nuclei, 
such as cluster structures and deformations. 
By using the AMD to describe the core structure, 
the DC wave function, therefore, can contain 
the cluster structure in the core nucleus as well as the core deformation.
Although the DC wave function is expected to be useful 
to analyze dineutron correlations in general nuclei, 
it may be too simple to describe 
details of realistic nuclear states because of the assumption of explicit dineutrons.
Thus, besides the DC wave function, we adopt the AMD wave function for the total system, 
and superpose the AMD wave function and the DC wave function 
to construct ``the AMD+DC wave function''. 
This AMD+DC method is a beyond AMD approach, 
which can efficiently supplement the AMD wave functions 
with dineutron correlation components.
Since the AMD+DC method can be applied to general nuclei without specific assumptions, 
it may be useful to investigate systematically 
dineutron correlations in various nuclei.

As a first step, we formulate the DC wave function and apply this framework to 
the system containing a 2$\alpha$ core and one dineutron, 
which corresponds to a toy model for the $^{10}$Be system. 
By calculating the energy expectation value of the dineutron with the DC wave function 
and analyzing its dependence on the dineutron size, 
we investigate the properties of a dineutron in the 2$\alpha$+2$n$ system and  
clarify the mechanism of the dineutron correlation enhancement at the nuclear surface. 
Then, superposing the 2$\alpha$+2$n$ DC wave functions and the $^{10}$Be AMD wave functions 
to construct the $^{10}$Be AMD+DC wave functions, 
we analyze the contribution of the dineutron components in $0^+$ states of $^{10}$Be. 
Effects of the core excitation on the dineutron correlation 
in $^{10}$Be($0^+$) are also discussed.

This paper is organaized as follows. 
In \S 2, we describe the framework of the new approach.
We explain the framework of the AMD wave function, the DC wave function, 
and the AMD+DC method, and describe the present method for the application to $^{10}$Be.
In \S 3, we refer to the effective Hamiltonian used in the present calculation.
In \S 4, we analyze the results of the 2$\alpha$+2$n$ DC wave function,
and in \S 5, we show the results of $^{10}$Be calculated with the AMD+DC method 
and discuss the dineutron components in $^{10}$Be.
In \S 6, a summary and an outlook are given.

\section{Framework}

We use two types of wave function to describe nuclear systems; 
the AMD wave function and the DC wave function.
At first, we explain the formalism of each wave function. 
Then, we describe the present DC wave function for the 2$\alpha$+2$n$ system, 
and explain the AMD+DC method for the $^{10}$Be system.
\\

\subsection{AMD wave function}
The AMD method has an advantage that the nuclear structure can be represented 
without specific assumptions such as axial symmetry and existence of clusters. 
In the AMD, the wave function $|\Phi_{\rm{AMD}} \rangle$ 
for a nuclear system with the mass number $A$ 
is described by a Slater determinant of $A$ single-particle wave functions as 
\begin{align}
|\Phi_{\rm{AMD}}(\boldsymbol{Z}) \rangle = &\
 \frac{1}{\sqrt{A!}} \det 
\big[ |\tilde{\psi}_i \rangle \big] 
\hspace{2em} \big( {\small \boldsymbol{Z} = \{ \boldsymbol{Z}_i \}
 = \{ \boldsymbol{Y}_i, \boldsymbol{\xi}_i \} \hspace{1em} (i=1, \cdots, A)} \big), \nonumber \\
= &\ \frac{1}{\sqrt{A!}} \det \big[ |\psi_{\boldsymbol{Y}_i} \rangle
                            |\chi_{\boldsymbol{\xi}_i} \rangle
                            |\tau_i \rangle \big] \label{eq:AMDwf}, \\
\mathrm{spatial \ part} : \ 
&\ \langle \boldsymbol{r}|\psi_{\boldsymbol{Y}_i} \rangle =
\left( \frac{2 \nu_i}{\pi} \right)^{\frac{3}{4}}
\exp[ - \nu_i ( \boldsymbol{r} - \boldsymbol{Y}_i )^2 ], \label{eq:AMDwf_space} \\
\mathrm{spin \ part} : \ 
&\ |\chi_{\boldsymbol{\xi}_i} \rangle = 
\xi_{i\uparrow} |\uparrow \rangle + \xi_{i\downarrow} |\downarrow \rangle, \\
\mathrm{isospin \ part} : \ 
&\ |\tau_i \rangle = |p \rangle \ \rm{or} \ |n \rangle,  
\end{align}
where the $i$-th single-particle wave function $|\tilde{\psi}_i \rangle$ is composed of 
the spatial part $|\psi_{\boldsymbol{Y}_i} \rangle$, 
the spin part $|\chi_{\boldsymbol{\xi}_i} \rangle$ and 
the isospin part $|\tau_i \rangle$. 
The subscripts of the spatial and spin parts, 
$\boldsymbol{Y}_i, \boldsymbol{\xi}_i$, 
represent the variational parameters for each single-particle wave function. 
The spatial part is approximated by a Gaussian wave packet 
which spreads around the center $\boldsymbol{Y}_i$
with the width $1/\sqrt{2\nu_i}$. 
The present AMD wave function is similar to 
that used in the fermionic molecular dynamics (FMD) 
\cite{feldmeier00,neff02}.
In principle, the width parameters can be different for each Gaussian wave packet
as Ref.~\citen{furutachi09}.
However, in the application to $^{10}$Be in the present work, 
we use the common width $\nu = 0.235$ fm$^{-2}$
for all single-particle wave functions for simplicity.

\subsection{DC wave function}
We here mention the general form of the DC wave function 
to deal with the dineutron condensate state consisting of a core and surrounding dineutrons. 
Later, we consider the specific case of one dineutron around the 2$\alpha$ core. 

In a DC wave function, 
a nucleus consists of a deformed core described with an AMD wave function
and some dineutrons around the core in the lowest $S$-orbit to form the condensate state. 
We formulate the general form of the DC wave function $|\Phi_{\rm DC} \rangle$, 
describing an $A$-nucleon system including $k_{0}$-dineutrons 
and the core with $A-2k_{0}$ nucleons as follows. 
\begin{align}
|\Phi_{\rm DC} \rangle = &\ 
\frac{1}{\sqrt{A!}} \prod_{k=1}^{k_{0}} \int d^3 \boldsymbol{Y}_{nk} 
\exp \left[ - \frac{Y_{nkx}^2}{B_{nx}^2} - \frac{Y_{nky}^2}{B_{ny}^2}
- \frac{Y_{nkz}^2}{B_{nz}^2} \right] \nonumber \\
&\ \times \det \big[ |\tilde{\psi}_1 \rangle \cdots |\tilde{\psi}_{A-2k_{0}} \rangle
|\tilde{\psi}_{n1 \uparrow} \rangle |\tilde{\psi}_{n1 \downarrow} \rangle \cdots
|\tilde{\psi}_{nk_{0} \uparrow} \rangle
|\tilde{\psi}_{nk_{0} \downarrow} \rangle \big], \label{eq:DCwf}
\end{align}
where the single-particle wave functions of the nucleons composing the core, 
$|\tilde{\psi}_i \rangle \ ( i = 1, \cdots, A-{2k_{0}} )$, 
are written by Gaussian wave packets
in the same form as those in the AMD wave function described in Eq.(\ref{eq:AMDwf}).
They are parametrized by the Gaussian centers and the spin orientations, 
$\{ \boldsymbol{Y}_i, \boldsymbol{\xi}_i \}$ $(i=1,\cdots,A-{2k_{0}})$.
The wave function for the $k$-th pair of neutrons, 
$|\tilde{\psi}_{nk \uparrow} \rangle |\tilde{\psi}_{nk \downarrow} \rangle
\ ( k = 1, \cdots, k_{0} )$ in the Slater determinant 
represents two neutrons with spin $\uparrow$ and $\downarrow$, 
whose spatial parts are expressed by Gaussian wave packets as
\begin{align}
|\tilde{\psi}_{nk \chi} \rangle \equiv 
|\psi_{nk} \rangle | \chi \rangle & |n \rangle, \\
\hspace{3em} \mathrm{spatial \ part} : &\
\langle \boldsymbol{r}|\psi_{nk} \rangle =
\left( \frac{2 \nu_{nk}}{\pi} \right)^{\frac{3}{4}}
\exp[ - \nu_{nk} ( \boldsymbol{r} - \boldsymbol{Y}_{nk} )^2 ], \label{eq:DCwf_space} \\
&\ \hspace{11.3em}
\nu_{nk} = \frac{1}{2b_{nk}^2},  \label{eq:dineutronwf} \nonumber \\
\mathrm{spin \ part} :  
&\ |\chi \rangle = | \uparrow \rangle \rm{or} | \downarrow \rangle. 
\end{align}
Two neutrons have the common spatial wave functions 
with the width parameter $\nu_{nk}$ and the Gaussian center $\boldsymbol{Y}_{nk}$, 
that is, they spread around the point $\boldsymbol{Y}_{nk}$ 
by an expansion $b_{nk}=1/\sqrt{2\nu_{nk}}$. 
We use the parameter $b_{nk}$ instead of $\nu_{nk}$ hereafter. 
By integrating the Slater determinants in the weight of the Gaussian distribution 
with respect to the centers of dineutrons $\boldsymbol{Y}_{nk}$, 
each dineutron spreads around the core in the $S$-orbit.
We can calculate analytically the Gaussian integral 
for the energy expectation value of the Hamiltonian given in \S3, 
and also the expectation values of the root mean square radius and density shown in \S5. 
The parameter $B_{n \sigma}$ in Eq.(\ref{eq:DCwf}) 
corresponds to the width of the dineutron distribution, 
and it is taken to be a common value for all dineutrons 
to describe a dineutron condensate state.
In principle, $B_{n \sigma}$ can be different values with respect to $\sigma=x,y,z$.
We, however, use the same value $B_n = B_{nx} = B_{ny} = B_{nz}$ 
in the present calculations for simplicity.
Then, by definition, the DC wave function contains dineutrons 
which correspond to the spin-zero neutron-neutron pairs 
with the relative orbital angular momentum $l=0$ and 
moving in the $S$-wave around the core. 
The general DC wave function defined above has
an analogy to the $\alpha$ condensate wave function proposed in Ref.~\citen{tohsaki01}
except for the existence of the core surrounded by dineutrons.
As stated later, we need to project the DC wave functions to the angular momentum eigenstates 
for the description of realistic states.
Note that the angular momentum projection of the DC wave functions is equivalent to 
that of the core wave functions, 
so the angular momentum projection can be achieved by
superposing the core wave functions.
We also should stress that the DC wave function is a microscopic $A$-body wave function 
with the full antisymmetrization between all nucleons including valence neutrons and core nucleons. 

\subsection{DC wave function for 2$\alpha$+2$n$}
Let us consider the particular case, $^{10}$Be (2$\alpha$+2$n$), 
where the system consists of two $\alpha$s and one dineutron 
(not a dineutron condensate state).
As indicated in previous studies, for instance Ref.~\citen{enyo99}, 
most states of $^{10}$Be have a rather developed 2$\alpha$ core, 
so we assume the core as 2$\alpha$. 
Below we explain the DC wave function for the 2$\alpha$+2$n$ system in detail. 
The 2$\alpha$+2$n$ DC wave function is written as
\begin{align}
|\Phi_{2\alpha + 2n} \rangle =  
\frac{1}{\sqrt{10!}} \int &\ d^3 \boldsymbol{Y}_{n} 
\exp \left[ - \sum_{\sigma=x,y,z}
\left( \frac{Y_{n \sigma}^2}{B_n^2} \right) \right] 
\det \big[ [|\tilde{\psi}_{\alpha 1} \rangle ]^4 [|\tilde{\psi}_{\alpha 2} \rangle]^4
|\tilde{\psi}_{n \uparrow} \rangle |\tilde{\psi}_{n \downarrow} \rangle \big], 
\end{align} 
where $|\tilde{\psi}_{\alpha} \rangle$ represents 
four wave functions composing an $\alpha$, that is, 
p$\uparrow$,  p$\downarrow$, n$\uparrow$ and n$\downarrow$. 
We can write the ten-body wave function in the Slater determinant 
with a product of the spatial and spin-isospin parts,
$\Phi^r_{2\alpha + 2n}$ and $\chi_{2\alpha + 2n}$,
and then we rewrite the $|\Phi_{2\alpha + 2n} \rangle$ as follows. 
\begin{align}
&|\Phi_{2\alpha + 2n} \rangle = \frac{1}{\sqrt{10!}} \mathcal{A} \left\{ \int d^3 \boldsymbol{Y}_{n} \exp \left[ - \sum_{\sigma}
 \left( \frac{Y_{n \sigma}^2}{B_n^2} \right) \right]
\Phi^r_{2\alpha + 2n}(d,\boldsymbol{Y}_n,b_n) \otimes \chi_{2\alpha + 2n} \right\}, \label{eq:1-DCwf} 
\\ \nonumber \\
&\Phi^r_{2\alpha + 2n}(d,\boldsymbol{Y}_n,b_n) \equiv
\left[ \psi_{\alpha 1} (\boldsymbol{r}_1) \cdots \psi_{\alpha 1} (\boldsymbol{r}_4) \ 
\psi_{\alpha 2} (\boldsymbol{r}_5) \cdots \psi_{\alpha 2} (\boldsymbol{r}_8) \ 
\psi_{n}(\boldsymbol{r}_9) \psi_{n}(\boldsymbol{r}_{10}) \right]
\label{eq:1-DCwf_space}
\end{align}
Here we fixed the distance between two $\alpha$s to $d$ and 
locate one $\alpha$ at $(0,0,d/2)$ and the other at $(0,0,-d/2)$.
For our purpose, we integrate $\boldsymbol{Y}_n$ with the Gaussian weight and 
rewrite two valence neutron wave functions in Eq.(\ref{eq:1-DCwf_space}) with 
the relative and center of mass wave functions of two neutrons in a dineutron, 
$\psi_{2n}$ and $\psi_G$. 
Then, the spatial part of Eq.(\ref{eq:1-DCwf}) can be rewritten as  
\begin{align}
\int d^3 \boldsymbol{Y}_n &\ \exp \left[ - \sum_{\sigma}
 \left( \frac{Y_{n \sigma}^2}{B_n^2} \right) \right] 
\Phi^r_{2\alpha + 2n}(d,\boldsymbol{Y}_n,b_n) \nonumber \\
 = &\ \frac{1}{\sqrt{10!}} 
\left[ \psi_{\alpha 1} (\boldsymbol{r}_1) \cdots \psi_{\alpha 1} (\boldsymbol{r}_4) \ 
\psi_{\alpha 2} (\boldsymbol{r}_5) \cdots \psi_{\alpha 2} (\boldsymbol{r}_8) \ 
\psi_{2n}(\boldsymbol{r}) \psi_G(\boldsymbol{r}_G) \right], \\
&\ \hspace{1em} \psi_{2n} \equiv \left( \frac{B_n}{b_n} \right)^3
\exp \left[ - \frac{r^2}{4 b_n^2} \right], 
\hspace{2em} \boldsymbol{r} = 
\boldsymbol{r}_9 - \boldsymbol{r}_{10}, \label{eq:DCwf_rel} \\
&\ \hspace{1em} \psi_{G} \equiv \exp \left[ - \frac{r_G^2}{\beta^2} \right], 
\hspace{6.2em} \boldsymbol{r}_G = 
\frac{\boldsymbol{r}_9 + \boldsymbol{r}_{10}}{2}, \hspace{1em}
\beta^2 \equiv B_n^2 + b_n^2 \label{eq:DCwf_com}. 
\end{align}
The coordinates $\boldsymbol{r}$ and $\boldsymbol{r}_G$ are 
the relative and center of mass coordinates for two neutrons.
As seen in Eq.(\ref{eq:DCwf_rel}) and Eq.(\ref{eq:DCwf_com}), 
the relative and center of mass wave functions are also written as the Gaussian forms
whose width parameters are $b_n$ and $\beta$, respectively, 
which satisfy the relation $b_n < \beta$ by definition.
It means that two neutrons spread by the expansion $b_n$ each other and 
the center of mass of the dineutron is distributed around the core by the extension $\beta$. 
So we consider the parameter $b_n$ as the size of the dineutron, 
and $\beta$ as the spread of the dineutron from the core, 
so that we characterize the behavior of the dineutron by $b_n$, 
and $\beta$ instead of $B_n$.
As a result, the 2$\alpha$+2$n$ DC wave function 
is specified by the parameters $d$, $b_n$, and $\beta$. 

The wave function $\psi_{2n} \psi_G$ indicates the two-neutron wave function 
in the spatial part of the DC wave function before the antisymmetrization.
In the case of $b_n \sim \beta$, the single-neutron wave function $\psi_n$ 
for two neutrons becomes a Gaussian located at the center of the total system ($B_n \sim 0$), 
and the two-neutron state corresponds to the uncorrelated limit of two valence neutrons 
moving around the core in the $s$-orbit with the width $b_n \sim \beta$. 
In contrast, when $b_n $ is relatively small compared with $\beta$, 
two neutrons form a compact dineutron moving in the $S$-wave around the core.

\subsection{AMD+DC wave function for $^{10}$Be}
In the DC wave function, all valence neutrons are assumed 
to couple to be spin-zero dineutrons.
This assumption is not necessarily appropriate for realistic nuclear systems. 
In the present study, to investigate dineutron components in $^{10}$Be, 
we construct the AMD+DC wave function 
by superposing the AMD wave function and the DC wave function defined above. 
We use the AMD wave functions for $^{10}$Be$(0^+_1)$ and $^{10}$Be$(0^+_2)$ 
obtained by the method of variation after parity and angular momentum projections (VAP).
The details of the VAP method in the AMD framework is described, 
for example, in Refs.~\citen{enyo98,enyo99}.
We consider the superposition of the states $|\Phi_{{\rm AMD}, k} \rangle$ 
given by the AMD wave functions $\Phi_{\rm AMD}(\boldsymbol{Z}^{(k)})$, 
and the states $|\Phi_{{\rm DC}, k} \rangle$ given by the DC wave functions 
having various sets of the parameters $(d_k,\beta_k,b_{nk})$; 
$|\Phi_{{\rm DC},k} \rangle = |\Phi_{\rm DC}(d_k,\beta_k,b_{nk}) \rangle$.
The numbers of the basis AMD wave functions and the DC wave functions are $k_{\rm AMD}$ and $k_{\rm DC}$, 
and $k$ is the label for the AMD and DC wave functions, i.e., $k=1,\cdots, k_{\rm AMD}$ for
$|\Phi_{{\rm AMD}, k} \rangle$ and $k=k_{\rm AMD}+1,\cdots, k_{\rm AMD}+k_{\rm DC}$ for 
$|\Phi_{{\rm DC}, k} \rangle$.
We project those $k_{\rm AMD}+k_{\rm DC}$ basis wave functions to the parity and angular momentum eigenstates 
with the projection operator 
$\mathcal{P}^{J \pm}_{MK}$ which projects to the eigenstate of
the total angular momentum $J$, the $z$-component of the angular momentum 
in the laboratory frame  $M$ and that of the body-fixed frame $K$, and the parity $\pm$. 
For $0^+$ states, the AMD+DC wave function is given 
by the superposition of the projected states as follows.
\begin{equation}
|\Psi_{\rm AMD+DC} \rangle \equiv
\sum_{k=1}^{k_{\rm AMD}} f_k \mathcal{P}^{0+}_{00}|\Phi_{{\rm AMD}, k} \rangle
+ \sum_{k=k_{\rm AMD}+1}^{k_{\rm AMD}+k_{\rm DC}} 
f_k \mathcal{P}^{0+}_{00}|\Phi_{{\rm DC}} 
(d_k, \beta_k, b_{nk} ) \rangle,  
\end{equation}
The coefficients $f_k$ are determined 
by diagonalization of the Hamiltonian and norm matrices  
by solving the Hill-Wheeler equation,
\cite{ring}
\begin{equation}
\mathcal{H} f = E \mathcal{N} f
\hspace{2em} 
\left\{ \begin{array}{l}
\mathcal{H}_{k k'} \equiv \langle \Phi_k|H| \Phi_{k'} \rangle, \\
\mathcal{N}_{k k'} \equiv \langle \Phi_k|\Phi_{k'} \rangle. 
\end{array} \right.
\end{equation}
We calculate the angular momentum projection numerically 
by replacing the integrals of Euler angles with the sum of minute meshes.

\section{Effective Hamiltonian}

The Hamiltonian used in the present calculation 
consists of the kinetic term and the effective nuclear forces,
\begin{equation}
H = T - T_G + V_{\rm cent} + V_{\rm LS} + V_{\rm Coul}, 
\end{equation}
where $T$ and $T_G$ are the total and center of mass kinetic energy. 
In our framework, it is difficult to exactly remove the center of mass motion
from the wave function 
because different width parameters are used in the DC wave functions. 
We, therefore, approximately treat the center of mass effect in the energy
by subtracting $T_G$ from $T$ in the Hamiltonian. 
The potential term $V_{\rm cent}$, $V_{\rm LS}$ and $V_{\rm Coul}$ are 
the central, spin-orbit and Coulomb force. 
We use the Volkov No.2 interaction as $V_{\rm cent}$ \cite{volkov65}, 
\begin{align}
V_{\rm cent} = &\ \sum_{k=1}^2 \left\{ \sum_{i<j}^A v_{k} 
\exp \left[ - \left( \frac{\boldsymbol{r}_i - \boldsymbol{r}_j}{a_k} \right)^2 \right] 
\right\} \times \mathcal{X}, \\
&\ \mathcal{X} \equiv W + B \mathcal{P}_{\sigma} 
- H \mathcal{P}_{\tau} - M \mathcal{P}_{\sigma} \mathcal{P}_{\tau}, \label{eq:cent_X}
\end{align}
where the operators $\mathcal{P}_{\sigma}$ and $\mathcal{P}_{\tau}$ are 
the spin- and isospin-interchanging operators.
We fixed the parameters as 
$v_1 = -60.65 \ {\rm MeV}, \ v_2 = 61.64 \ {\rm MeV}, \ 
a_1 = 1.80 \ {\rm fm}, \ a_2 = 1.01 \ {\rm fm}$.
As for the spin-orbit force, we use the G3RS interaction
\cite{tamagaki68,yamaguchi79},
\begin{equation}
V_{\rm LS} = \sum_{k=1}^2 \left\{ \sum_{i<j}^A v_k 
\exp \left[ - \left( \frac{\boldsymbol{r}_i - \boldsymbol{r}_j}{a_k} \right)^2 \right] 
\mathcal{P}(^3 O) \ \boldsymbol{L}_{ij} \cdot \boldsymbol{S}_{ij} \right\}, \label{eq:G3RS} 
\end{equation}
where $\boldsymbol{L}_{ij}$ and $\boldsymbol{S}_{ij}$ are the relative orbital angular momentum 
and the total spin of two particles, and the parameters are
$a_1 = 0.447 \ {\rm fm}, \ a_2 = 0.6 \ {\rm fm}$. 
The operator $\mathcal{P}(^3 O)$ projects two particles to the triplet-odd state. 
The Coulomb force $V_{\rm Coul}$ is approximated by a sum of seven Gaussians. 

In the present calculation, we adopt the same interaction parameters as those used
in Ref.~\citen{suhara10}, that is, $B = H = 0.125, \ M = 0.60$ in Eq.(\ref{eq:cent_X}), and 
$v_1 = - v_2 = 1600 \ {\rm MeV}$ in Eq.(\ref{eq:G3RS}).

\section{A dineutron around a 2$\alpha$ core}

As the first step, we apply the DC wave function to the 2$\alpha$+2$n$ system
to understand the properties of a dineutron around the 2$\alpha$ core. 
We discuss the mechanism of the dineutron correlation enhancement 
by analyzing the energy of the dineutron as a function of the dineutron size $b_n$ 
and the spread of the dineutron distribution $\beta$.

We define the dineutron energy, $E_{2n}$, measured from the limit 
where the $2\alpha$ core exists at the origin and two neutrons are free, 
i.e. in the zero-momentum state, 
\begin{align}
E_{2n}(d, \beta, b_n) = &\ E_{2\alpha+2n}(d, \beta, b_n) - E_{2\alpha}(d), \\ 
\nonumber \\
E_{2\alpha+2n}(d,\beta,b_n) \equiv &\ 
\frac{\langle \Phi_{2\alpha+2n}(d, \beta, b_n)|H|\Phi_{2\alpha+2n}(d, \beta, b_n) \rangle}
{\langle \Phi_{2\alpha+2n}(d, \beta, b_n)|\Phi_{2\alpha+2n}(d, \beta, b_n) \rangle}, \\
E_{2\alpha}(d) \equiv &\
E_{2\alpha+2n}(d,\beta,b_n)|_{\beta \rightarrow \infty, b_n \rightarrow \infty}, 
\end{align}
where $E_{2\alpha}(d)$ is the energy expectation value for the DC wave function 
in the infinite $b_n$ and $\beta$ limit. 
With a fixed $d$ for the 2$\alpha$ distance and 
a fixed $\beta$ for the expansion from the core, 
the energy $E_{2n}(d, \beta, b_n)$ is regarded 
as the binding energy of two neutrons virtually confined around the $2\alpha$ core.

\begin{figure}[t]
\begin{center}
\begin{tabular}{cc}
{\includegraphics[scale=0.6]{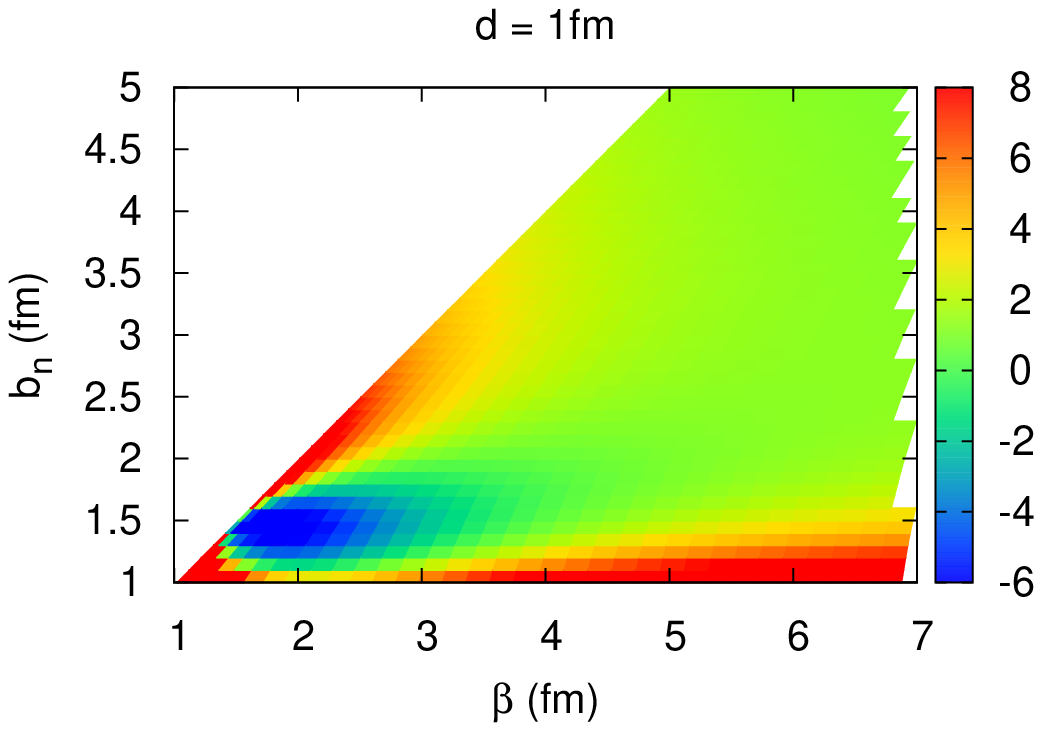}} &
\hspace{-14mm}
{\includegraphics[scale=0.6]{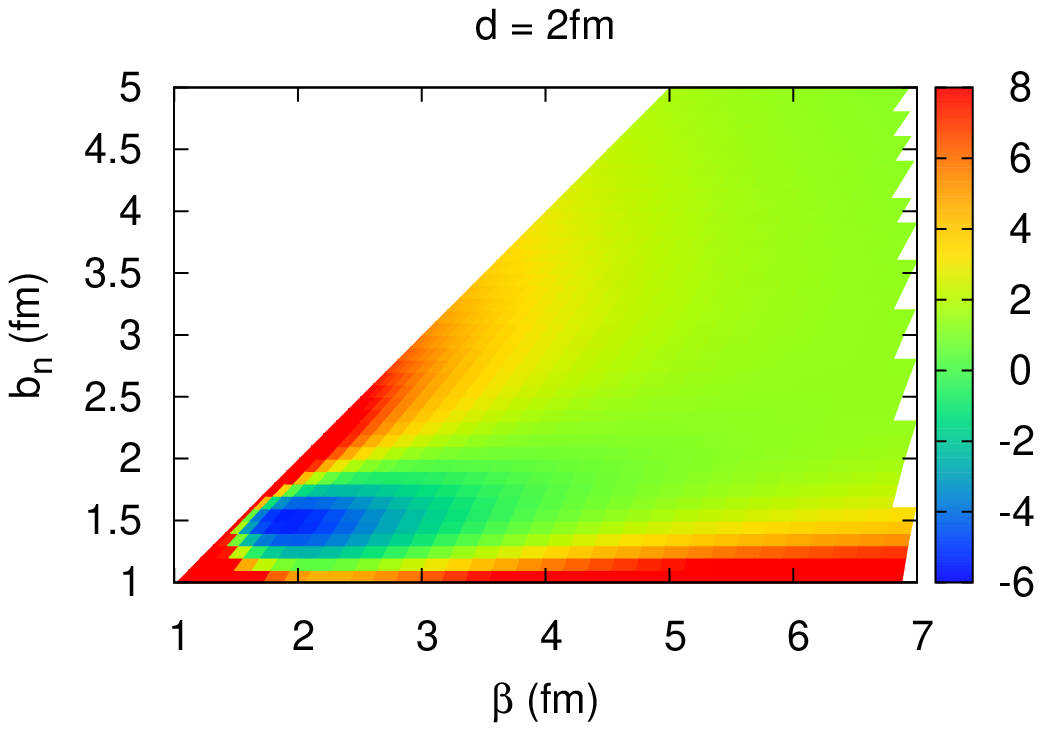}} \vspace{-5mm} \\
{\includegraphics[scale=0.6]{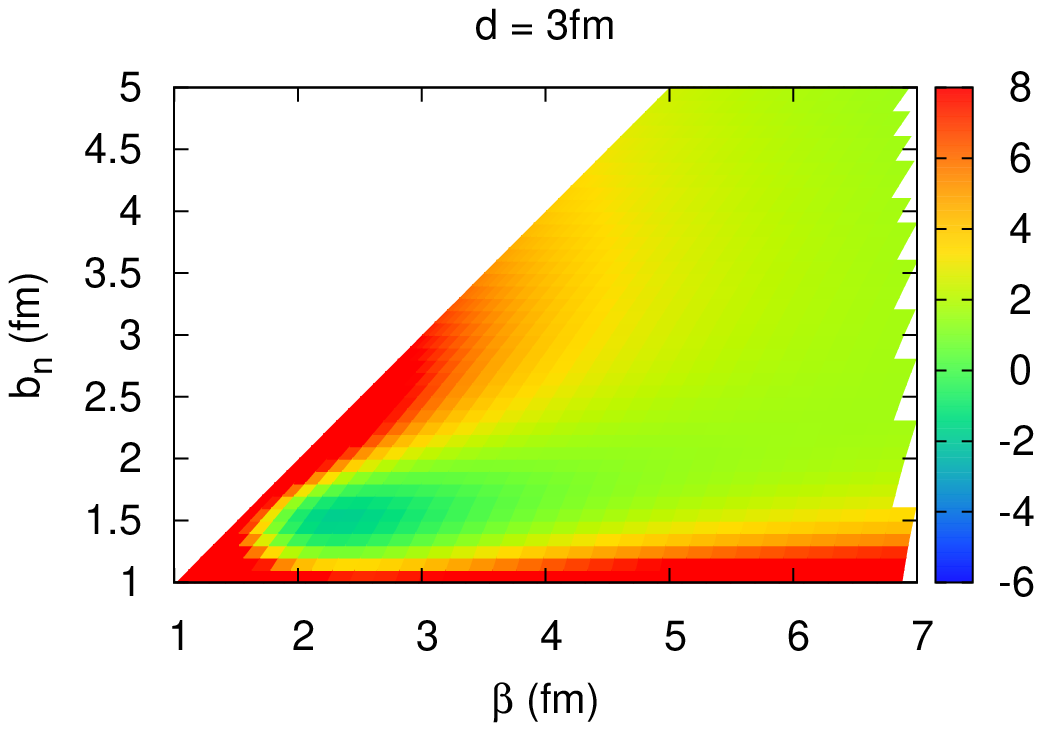}} &
\hspace{-13mm}
{\includegraphics[scale=0.6]{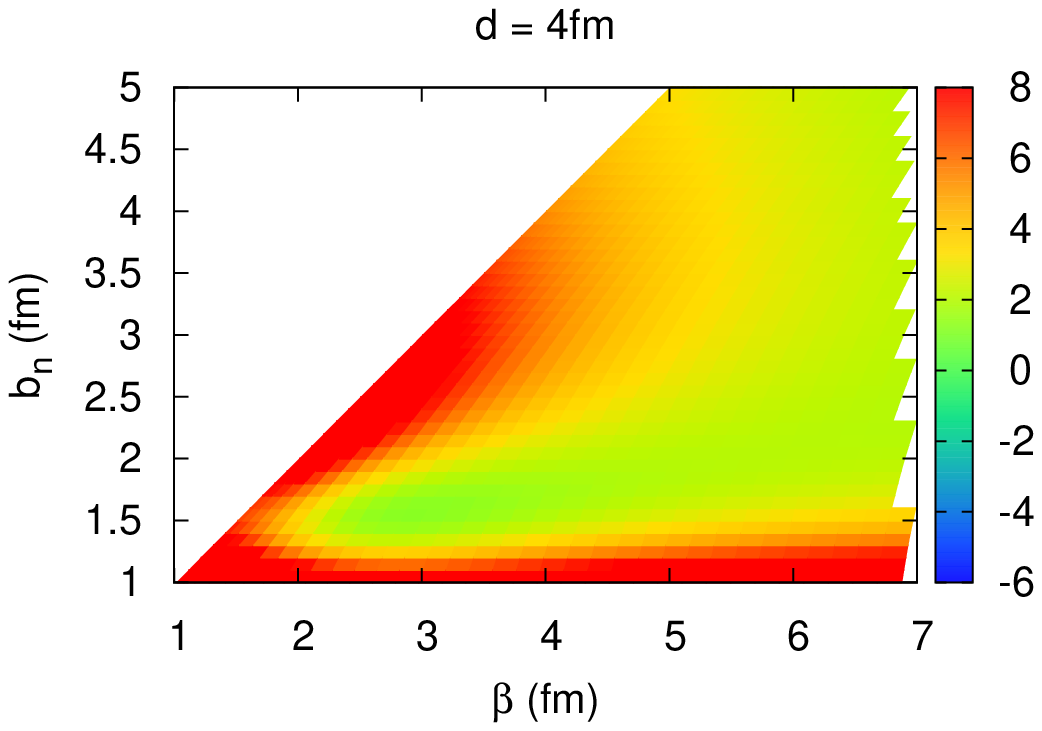}} \\
\end{tabular}
\vspace{-5mm}
\caption{\footnotesize 
Energy surfaces on the $\beta$-$b_n$ plane 
of the dineutron energy $E_{2n}$ in the 2$\alpha$+2n system. 
The fixed distances of two $\alpha$s, $d$, are shown above each figure. 
One must pay attention to the figure restricted by $b_n < \beta$ 
because of the definition of $\beta$ 
(see Eq.(\ref{eq:DCwf_com})).}
\label{fig:E_B_b}
\end{center}
\end{figure}
 
At first, we consider the behavior of the dineutron energy $E_{2n}$ 
when its size and expansion from the core, $b_n$ and $\beta$, are changed. 
Here we fixed the $\alpha$-$\alpha$ distance $d$ to 1, 2, 3, 4 fm.
We show in Fig.~\ref{fig:E_B_b} the dineutron energy $E_{2n}$ 
plotted as functions of $(d,\beta,b_n)$.
Since the qualitative feature does not depend on the parameter $d$ so much, 
we focus only on the case $d=2$~fm (the upper-right figure in Fig.~\ref{fig:E_B_b}), 
which gives the lowest solution 
for the energy $E_{2\alpha + 2n}$ of the total 2$\alpha$+2n system. 
The energy minimum exists at $(\beta, b_n) \sim (1.9, 1.5)$. 
As the viewpoint of the dependence on the dineutron size $b_n$, 
the energy increases rapidly when $b_n$ deviates from the minimum point $b_n \sim 1.5$~fm. 
This means that a dineutron around the 2$\alpha$ core favors a fairly compact size. 
On the other hand, the energy increases only gradually 
when the spread from the core $\beta$ increases from $\beta \sim 1.9$~fm. 
Thus, the dineutron can extend far from the 2$\alpha$ core to some extent. 
These results suggest that the dineutron around the 2$\alpha$ core tends to 
be distributed widely keeping a compact size. 

\begin{figure}[t]
\begin{minipage}{0.4\hsize}
\begin{center}
\begin{subfigure}
{\vspace{-7mm}
\hspace{2mm}
\includegraphics[scale=0.7]{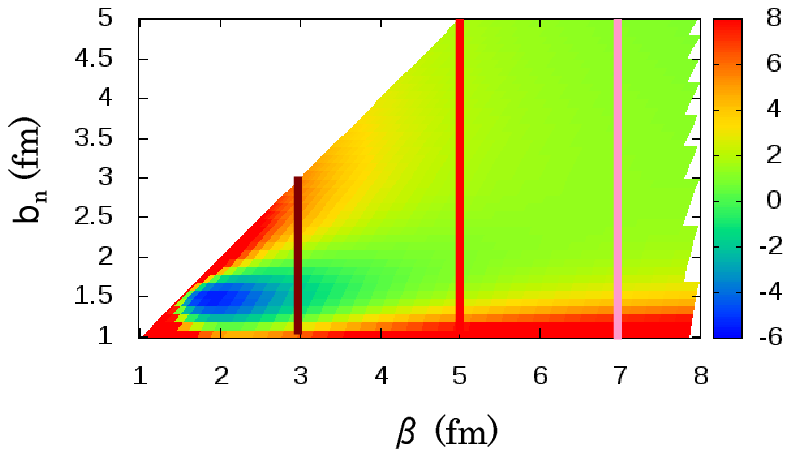}}
\end{subfigure}
\end{center}
\end{minipage}
\begin{minipage}{0.6\hsize}
\begin{center}
\begin{subfigure}
{\hspace{5mm}
\includegraphics[scale=0.6]{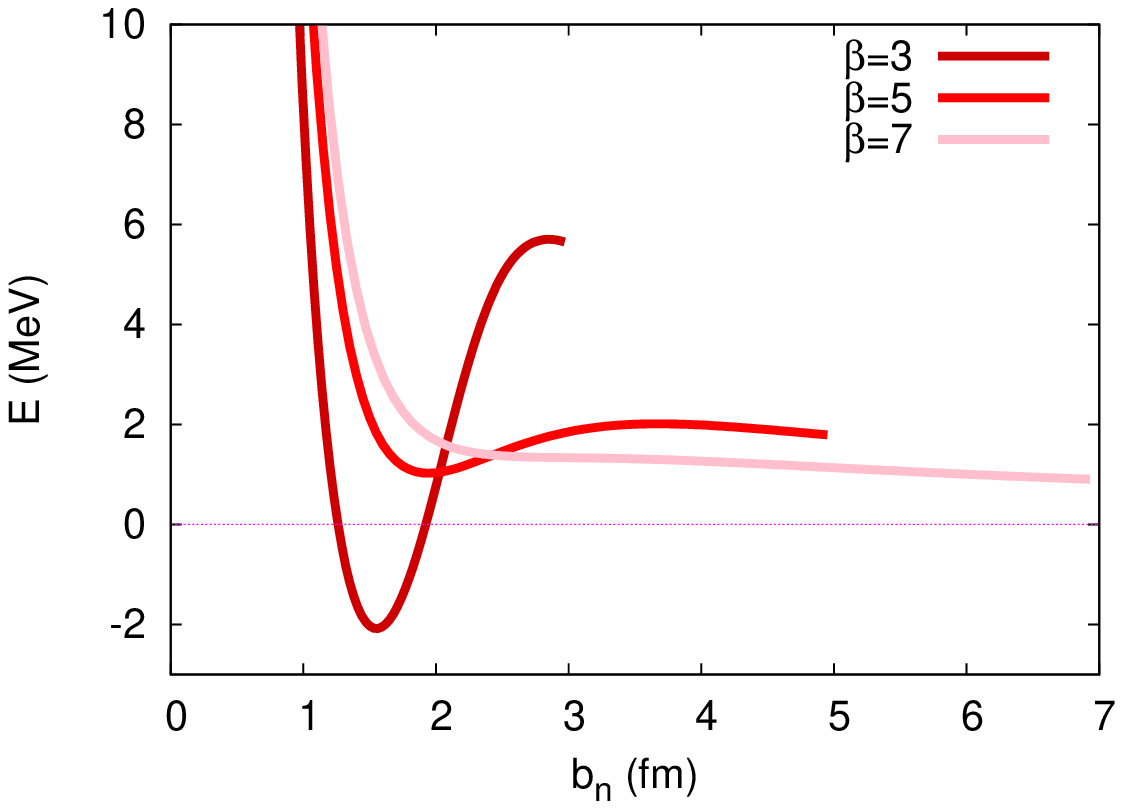}} \\
\end{subfigure}
\end{center}
\end{minipage}
\vspace{-7mm}
\caption{\footnotesize
The dineutron energy, $E_{2n}$, in the 2$\alpha$+2$n$ system 
for the fixed $2\alpha$ distance $d=2$~fm. 
The left figure is the energy surface on the $\beta$-$b_n$ plane 
same as the figure of $d=2$~fm in Fig.~\ref{fig:E_B_b}. 
The right one is $E_{2n}$ plotted as a function of $b_n$ 
for the fixed $\beta = 3, 5, 7$~fm, 
which corresponds to the section on the line 
$\beta = 3, 5, 7$ fm of the energy surface shown in the left.}
\label{fig:E_b_2fm_357}
\end{figure}

Subsequently, we discuss the origin of the compact dineutron formation below. 
We analyze the $b_n$-dependence of the dineutron energy 
with a fixed $\beta$-value in more detail. 
The DC wave function with a fixed $\beta$ corresponds to the state
where a size-chageable dineutron is confined to a finite region around the core. 
In Fig.~\ref{fig:E_b_2fm_357}, 
we plot the dineutron energy as a function of the dineutron size $b_n$ 
with $\beta$ fixed to be 3, 5, 7~fm. 
When $\beta = 3$~fm, that is, the dineturon is confined relatively close to the core, 
in other words, near the surface of the nucleus, 
the energy structure shows a high barrier at the point $b_n \sim \beta$ 
and a deep pocket at $b_n \sim 1.5$~fm. 
Such structure disappears gradually as the spread from the core, $\beta$, increases. 
The state $b_n \sim \beta$ corresponds to the uncorrelated limit of two valence neutrons, 
while the state $b_n \sim 1.5$~fm corresponds to the state 
where two neutrons correlate strongly to some extent to form a compact dineutron. 
When two valence neutrons are distributed with $\beta = 3$ fm, 
i.e. near the nuclear surface, 
they are strongly correlated to be a rather compact size due to the pocket in their energy, 
and the high barrier prevents two neutrons from separating each other by a large distance. 

\begin{figure}[t]
\begin{center}
\begin{tabular}{cc}
{\includegraphics[scale=0.55]{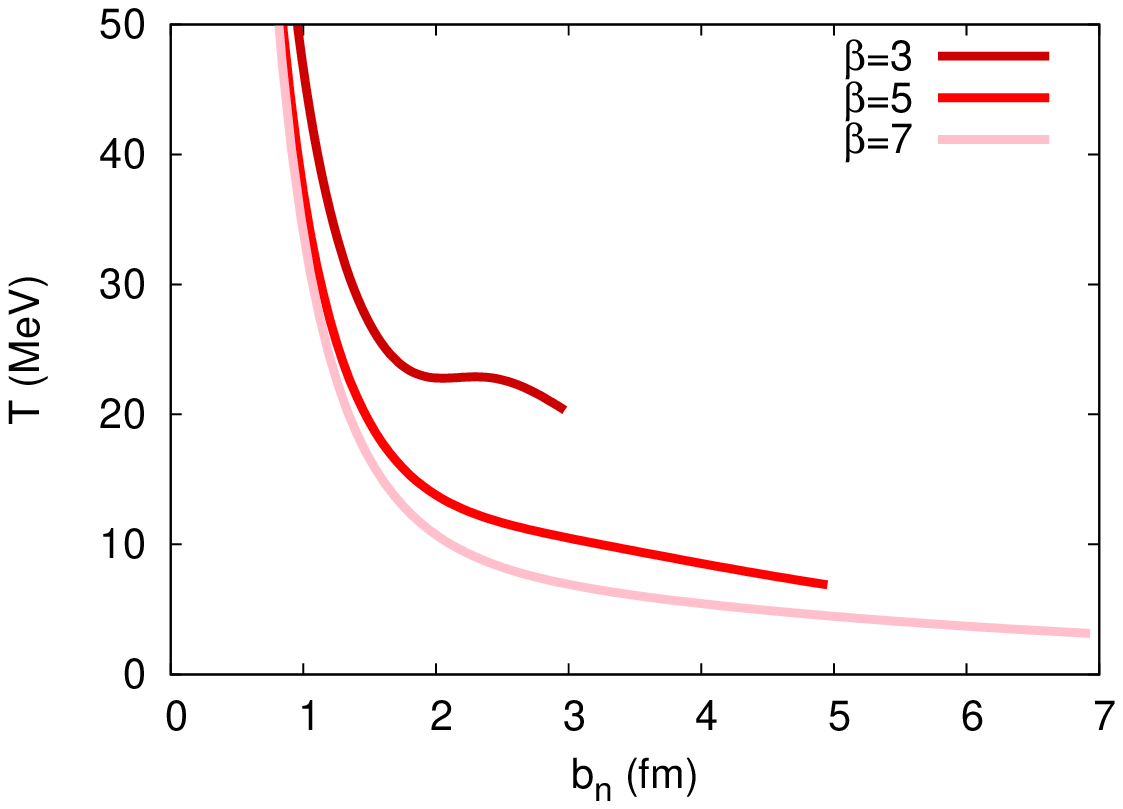}} & 
\hspace{-10mm}
{\includegraphics[scale=0.55]{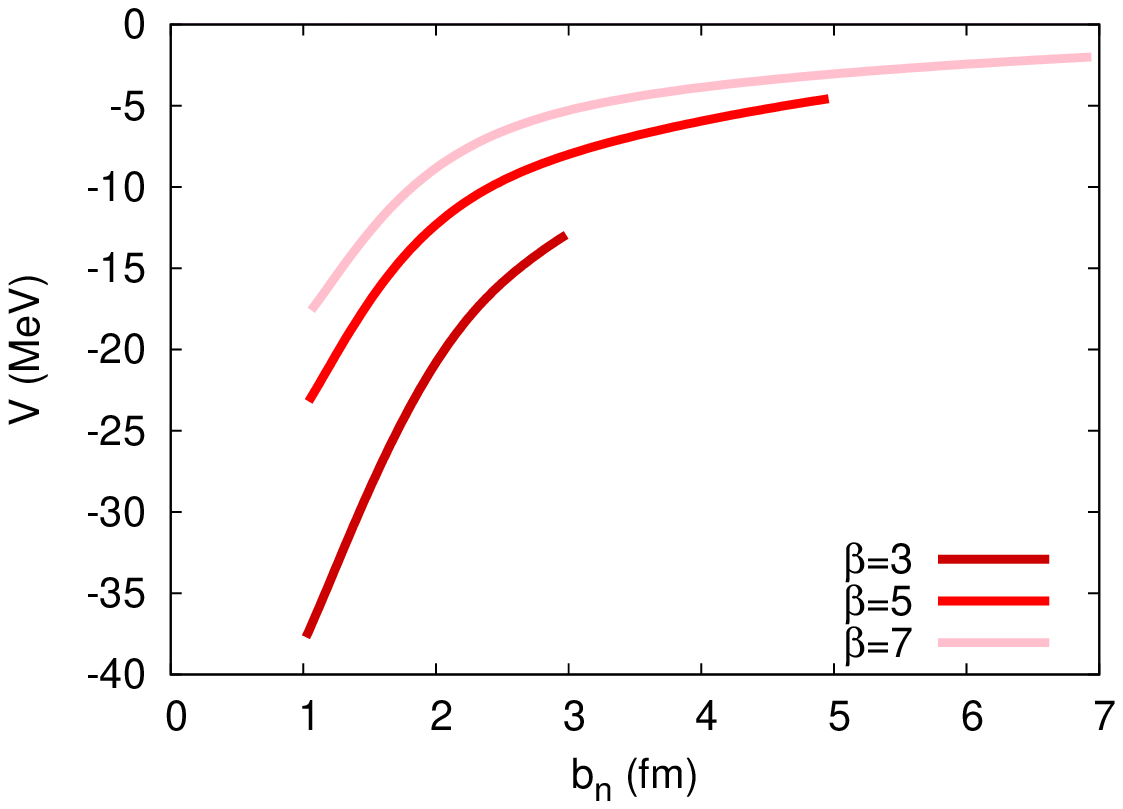}} \\
\end{tabular}
\vspace{-3mm}
\caption{\footnotesize
The left and the right figure shows the kinetic and the potential energy of the dineutron 
around the 2$\alpha$ core, respectively. }
\label{fig:T_V_b_2fm}
\end{center}
\end{figure}

As mentioned above, the existence of the barrier is essential to enhance the dineutron correlation.
This barrier structure suggests that 
two neutrons in the uncorrelated limit feel the effective repulsion.
In order to clarify this mechanism of developing the dineutron, 
we decompose the energy $E_{2n}$ 
into the kinetic part and the potential part (Fig.~\ref{fig:T_V_b_2fm}). 
The potential energy decreases monotonously as $\beta$ or $b_n$ increases, 
as can be interpreted easily. 
As the dineutron size $b_n$ becomes small, 
two neutrons gain the potential energy due to 
the attraction of the neutron-neutron interaction. 
The decrease of $\beta$ corresponds to 
the shrinkage of the confining region of the dineutron, 
and therefore the attraction from the core increases. 
This simple $b_n$-dependence of the potential term does not contribute to 
the barrier structure of the dineutron energy $E_{2n}$.

On the other hand, the kinetic energy has a distinctive structure. 
The kinetic part of the dineutron energy for $\beta = 3$~fm, 
where the dineutron are distributed near the nuclear surface, 
shows the plateau at the region $b_n \sim \beta$,  
while that for $\beta = 5, 7$~fm decreases monotonously as $b_n$ increases. 
This monotonous decrease is due to the uncertainty principle 
for two neutrons in the dineutron. 
As the size of a dineutron $b_n$ becomes smaller, 
the uncertainty principle enlarges the kinetic energy of two neutrons in it, 
and vice versa. 
When the dineutron is distributed near the surface of the nucleus, 
in addition to such an effect, 
another effect arises from the Pauli blocking against the core, and it is the main origin of the plateau. 
When the dineutron is distributed near the core, 
the Pauli exclusion principle works significantly 
between the neutrons in the dineturon and those in the core. 
The state at $b_n \sim \beta$ corresponds to the uncorrelated limit state 
where each valence neutron occupies the isotropic $s$-orbit 
with the width $b_n$ around the core. 
Since the neutrons in the core already occupy the $s$-orbit with the width $1/\sqrt{2\nu}$, 
two valence neutrons feel the strong Pauli repulsion from the core neutrons.
The increase of the Pauli repulsion constructs the plateau 
at the region $b_n \sim \beta$ in the kinetic term.
Whereas, when the dineutron size $b_n$ is small relatively to $\beta$, 
two neutrons have the strong angular correlation 
and their single-particle wave functions contain components of higher angular momentum orbits. 
Then the Pauli blocking effect from the core becomes weak, 
and the Pauli repulsion in the kinetic term is suppressed. 
In this way, the plateau of the kinetic part of the dineutron energy for $\beta = 3$~fm 
originates in the Pauli repulsion from the core. 
In cases of $\beta = 5$ and 7~fm, since the dineutron distributes far from the core 
and the Pauli repulsion is generally weak, 
there is no clear structure of such a plateau in the kinetic energy.

In the above analysis, it is concluded that
the high barrier and the deep pocket seen 
in the dineutron energy $E_{2n}$ for $\beta = 3$~fm 
are constructed by the plateau structure of the kinetic part 
and the attraction of the potential part. 
The effects from the core are essential for these structures, 
i.e. the Pauli repulsion from the core yields the barrier, 
and the additional attraction from the core contributes to the deep pocket. 
Because of the barrier and pocket structure in the dineutron energy, 
a dineutron at the nuclear surface favors a suitable compact size.
Also in the $\alpha$+2$n$ and 3$\alpha$+2$n$, 
we find a similar mechanism for dineutron formation, 
though the depth of the energy pocket depends on the core. 
Since the dineutron size at the energy pocket is common for these nuclei, 
it turns out that the dineutron correlation occurs in the circumstance 
of the $s$-orbit occupied by core neutrons and the unoccupied $p$-orbits.
This mechanism of the dineutron correlation enhancement 
due to the Pauli repulsion from the core might occur in general nuclei.
The favored size, however, may depend on the allowed orbits for each nucleus.

\section{Contribution of a dineutron component to $^{10}$Be states}

\subsection{Superposition of AMD and DC wave functions}
In this section, we apply the AMD+DC method to $^{10}$Be 
and analyze the dineutron component in $^{10}$Be states.
Since, here, each DC wave function has an isotropic parameter $B_n$ ($= B_{nx} = B_{ny} = B_{nz}$), 
we discuss only $0^+$ states which are thought to be described properly 
in this parameter fixing. 
We adopt the AMD wave functions corresponding to the ground state and 
the first $0^+$ excited state obtained by the VAP. 
We superpose DC wave functions having various $b_n$- and $\beta$-values, 
with two AMD wave functions of $^{10}$Be($0^+_1, 0^+_2$). 
We superpose DC wave functions not only with $d=2$ fm, 
but also having different $d$-values. 
The superposition of DC wave functions with  different $d$-values
corresponds to the situation where two $\alpha$s can move relatively 
and, consequently, the consideration of the core excitation.  
We also perform the fixed-$d$ calculations 
and compare the results with the superposing-$d$ calculations 
to see effects of the core excitation.
Thus, we consider the following superpositions,
\begin{align}
&{\rm (i)}\ \Psi = \sum_{k=1}^2 f_k \mathcal{P}^{0+}_{00} \Phi_{{\rm AMD}, k}, \\ 
&{\rm (ii)}\ \Psi = \sum_{k=3}^{18} f_k \mathcal{P}^{0+}_{00} 
\Phi_{{\rm DC}}(d=2,\beta_k,b_{nk}), \\
&{\rm (iii)}\ \Psi = \sum_{k=1}^2 f_k \mathcal{P}^{0+}_{00} \Phi_{{\rm AMD}, k} +
\sum_{k=3}^{18} f_k \mathcal{P}^{0+}_{00} \Phi_{{\rm DC}}(d=2,\beta_k,b_{nk}), \\
&{\rm (iv)}\ \Psi = \sum_{k=1}^2 f_k \mathcal{P}^{0+}_{00} \Phi_{{\rm AMD}, k} +
\sum_{k=3}^{66} f_k \mathcal{P}^{0+}_{00} \Phi_{{\rm DC}}(d_k,\beta_k,b_{nk}).
\end{align}
For the parameters of basis DC wave functions, 
we adopt $\beta = 2, 3, 4, 5$ fm, 
and four $b_n$-values for each $\beta$ in the following way \cite{hiyama03}, 
\begin{equation}
b_{ni} = b_{n1} \times \left( \frac{b_{ni_{\rm max}}}{b_{n1}} \right)
^{(i-1)/(i_{\rm max}-1) } 
\hspace{2em} (i=1, \cdots i_{\rm max}), \label{eq:b_n_choise}
\end{equation}
where we choose $i_{\rm max} = 4$, 
$b_{n1} = 1.5$ fm and $b_{ni_{\rm max}} = \beta - 0.2$ fm. 
The parameter $d$ is chosen to 1, 2, 3, 4 fm, 
when superposing different-$d$ DC wave functions ((iv)),
so that the number of the basis DC wave functions in the superposition 
is $k_{\rm DC}=16$ for (ii), (iii), and $k_{\rm DC}=64$ for (iv). 
We hereafter label the wave functions (i) AMD, (ii) DC, 
(iii) AMD+DC($d=2$) and (iv) AMD+DC($\Sigma_d$). 

\subsection{Energy of $0^+$ states in $^{10}$Be}
\begin{figure}[b]
\begin{center}
\includegraphics{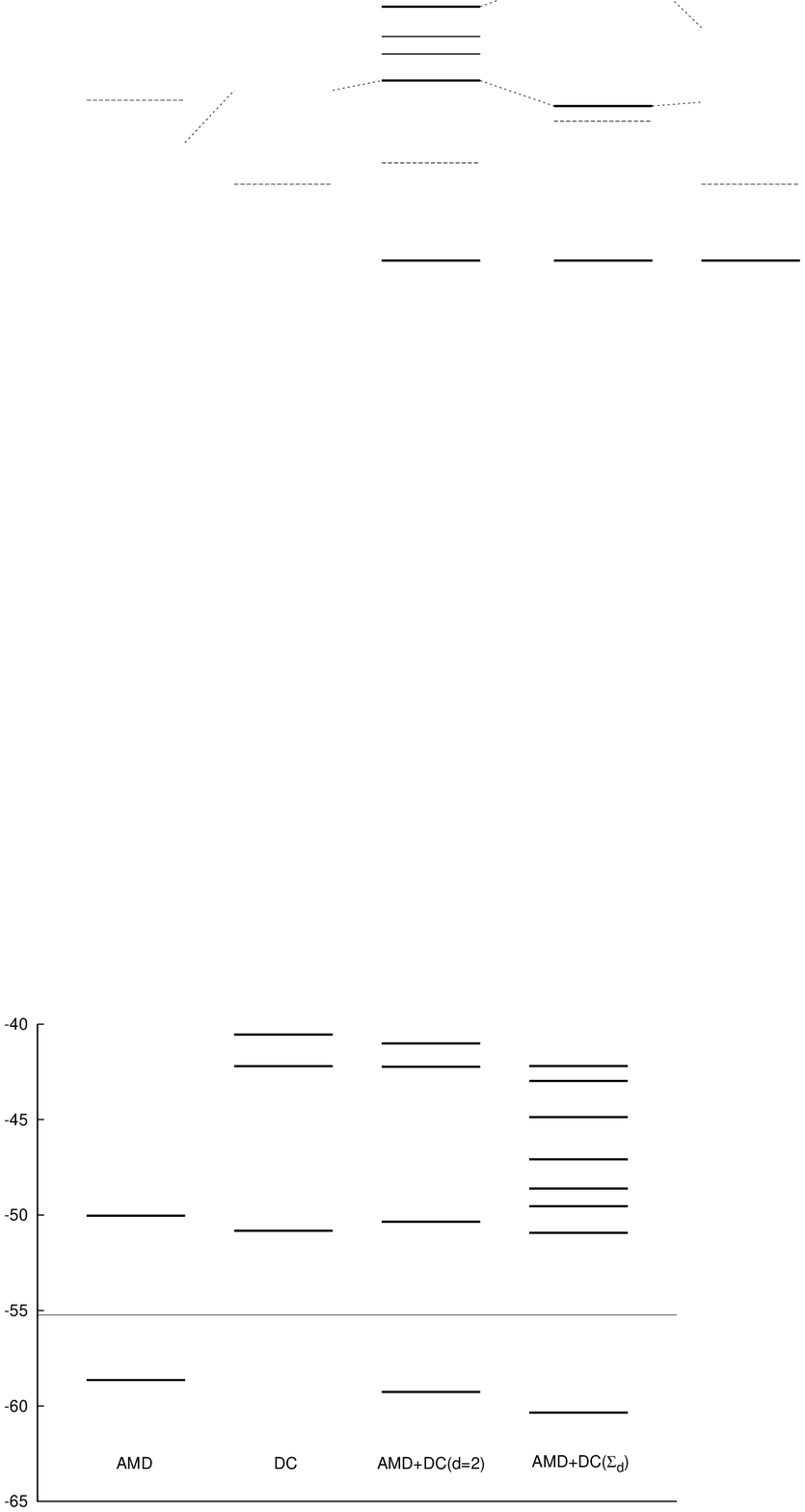}
\caption{\footnotesize
The energy spectra of $0^+$ states of $^{10}$Be. 
The labels AMD, DC, AMD+DC($d=2$) and AMD+DC($\Sigma_d$) indicate the spectrum 
calculated with only AMD wave functions (i), 
the one with only DC wave functions fixed $d$ to 2 fm (ii), 
the one with the AMD+DC wave function whose parameter $d$ is fixed to 2 fm (iii) 
and the AMD+DC wave function whose parameter $d$ is superposed (iv), respectively. 
The calculated threshold to $\alpha$+$\alpha$+$n$+$n$ ($-55.23$ MeV) is also shown.}
\label{fig:Be10_energy_spectrum}
\end{center}
\end{figure}
First of all, we show the energy spectra 
of AMD, DC, AMD+DC($d=2$) and AMD+DC($\Sigma_d$)
for comparison in Fig.~\ref{fig:Be10_energy_spectrum}. 
The lowest and second lowest levels in each calculation, 
(i), (iii) and (iv), correspond to the $0^+_1$ and $0^+_2$ states 
and their energies are $-58.64$, $-50.03$ (AMD), $-59.27$, $-50.34$ (AMD+DC($d=2$)), 
and $-60.35$, $-50.93$ (AMD+DC($\Sigma_d$)) in MeV, respectively. 
In addition to the $0^+_1$ and $0^+_2$ states, 
many states are obtained by the diagonalization of the AMD+DC 
in the higher energy region. 
Although most of them show features of unbound continuum states, 
we obtain an indication of the possible candidate for a developed dineutron state 
in the fifth state in AMD+DC($\Sigma_d$).
We will refer to that state in the end of this section. 

In the AMD wave functions describing the $0^+_1$ and $0^+_2$,  
the energy of the $0^+_2$ is higher than that of the $0^+_1$ by 8.61 MeV 
(6.18 MeV for an experimental value). 
In studies of $^{10}$Be such as Refs.~\citen{enyo99} and \citen{itagaki00}, 
it is indicated that the $0^+_1$ state has the 2$\alpha$ 
surrounded by two neutrons  distributed
predominantly in the $\pi^2$ configuration in terms of the molecular orbit,  
while the $0^+_2$ state has a remarkably developed 2$\alpha$ core structure 
with two neutrons in the $\sigma^2$ configuration.

Let us consider the energy gain in the $0^+_1$ and $0^+_2$ 
contributed by the DC wave functions. 
Comparing the energy between AMD, AMD+DC($d=2$) and AMD+DC($\Sigma_d$), 
we find that some energy is gained by mixing DC wave functions.
The energy gain in AMD+DC($\Sigma_d$) ($\sim 1$ MeV) is larger than that 
in AMD+DC($d=2$) in both states.
It indicates that the core excitation significantly contributes to 
at least the energy of those states. 
Not only the energy but also the dineutron component 
is affected by the core excitation as shown later. 
These results indicate that, 
in mixing the 2$\alpha$+2$n$ DC wave functions, 
the variation in the $\alpha$-$\alpha$ distance is important, 
and generally, the explicit core structure, such as the core deformation or polarization, 
may enhance the dineutron component. 

\begin{figure}[b]
\begin{center}
\includegraphics{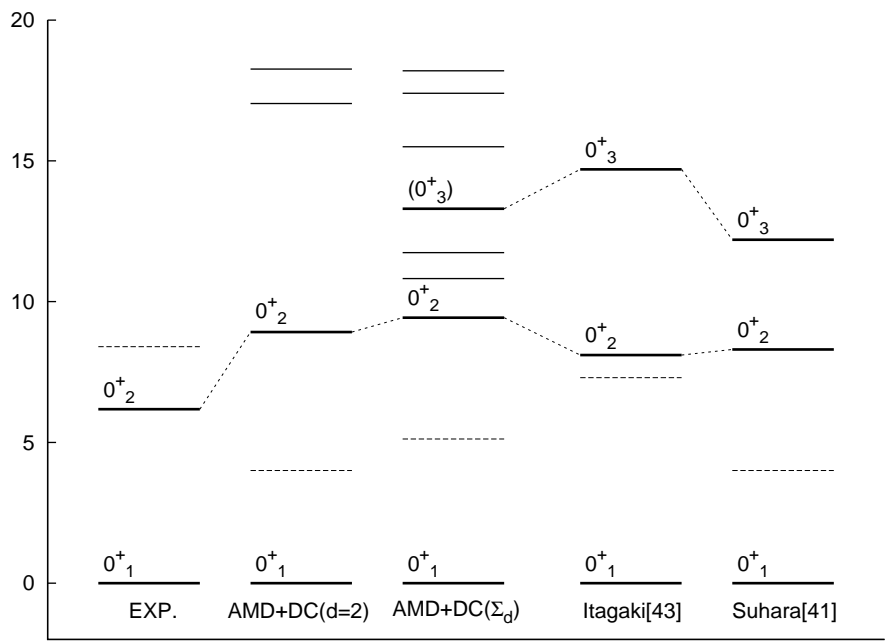}
\caption{\footnotesize
The energy spectra of $0^+$ states of $^{10}$Be 
on the basis of the ground state $0^+_1$ energy. 
The one labeled with Itagaki[43] is the calculated values with 
the molecular orbital model in Ref.~\citen{itagaki00}, and Suhara[41] is the values 
with the $\beta$-$\gamma$ constrained AMD in Ref.~\citen{suhara10}. 
The broken lines correspond to the threshold to $\alpha$+$\alpha$+$n$+$n$
for the experiment and each calculation.}
\label{fig:Be10_excited_energy_spectrum}
\end{center}
\end{figure}

We compare our results with the energy spectra of other calculations 
in Refs.~\citen{itagaki00,suhara10} 
as well as the experimental ones
in Fig.~\ref{fig:Be10_excited_energy_spectrum}. 
We show the spectra calculated with the molecular orbital (MO) model 
in Ref.~\citen{itagaki00}
and those with the $\beta$-$\gamma$ constraint AMD in Ref.~\citen{itagaki00}. 
In both works, 
the same effective interactions were adopted. 
The only difference in the used interactions is the strength of the spin-orbit force,
the parameters $v_1 = - v_2 = 1600$ MeV were used in 
Ref.~\citen{suhara10} as well as the present calculation, 
while $v_1 = - v_2 = 2000$ MeV were adopted in Ref.~\citen{itagaki00}.
Both works suggested the $0^+_3$ state in addition to the $0^+_1$ and $0^+_2$ states.
The present results of the $0^+_2$ excitation energy 
and the position of the 2$\alpha$+2$n$ threshold energy are
consistent with those of Ref.~\citen{suhara10} by Suhara {\it et al}. 
They underestimate the relative position of
the $2\alpha+2n$ threshold energy to the $0^+_1$ and $0^+_2$ states
compared with the experimental data and also with the work in Ref.~\citen{itagaki00}.
The main reason for the difference in the threshold position between 
the present and Itagaki {\it et al}.'s calculations is 
the difference in the spin-orbit interaction parameter. 
When we use the same interaction parameters as those used in Ref.~\citen{itagaki00}, 
we obtain the better result of the threshold position. 
For the quantitative reproduction of the energy levels as well as the threshold energy, 
we need fine tuning of the effective interactions.

\subsection{Radii and density for proton and neutorn distributions}
\begin{table}[t]
\begin{center}
\caption{\footnotesize
The root mean square radii of matter, proton and neutron 
of $^{10}$Be ($0^+_1$,$0^+_2$) 
calculated with the AMD, AMD+DC($d = 2$) and AMD+DC($\Sigma_d$) wave functions. 
The labels are the same as that in Fig.~\ref{fig:Be10_energy_spectrum}. 
The experimental value of the matter radius of the $0^+_1$ state is also shown.}
\label{tab:Be10_r_rms}
\begin{tabular}{cccccc}\hline \hline
& & \hspace{2em} AMD \hspace{2em} & AMD+DC($d=2$) & AMD+DC($\Sigma_d$) & EXP. \\ \hline
$0^+_1$ & $r_{\rm rms}$ & 2.28 & 2.24 & 2.32 & 2.30 $\pm$ 0.02 \\
& $r_{\rm rms}^p$ & 2.22 & 2.22 & 2.31 & \\
& $r_{\rm rms}^n$ & 2.32 & 2.34 & 2.41 & \\ \hline
$0^+_2$ & $r_{\rm rms}$ & 2.84 & 2.84 & 2.94 & \\
& $r_{\rm rms}^p$ & 2.72 & 2.68 & 2.66 & \\
& $r_{\rm rms}^n$ & 2.93 & 2.93 & 3.14 & \\ \hline \hline
\end{tabular}
\end{center}
\end{table}

\begin{figure}[htbp]
\begin{center}
{\includegraphics[scale=0.55]{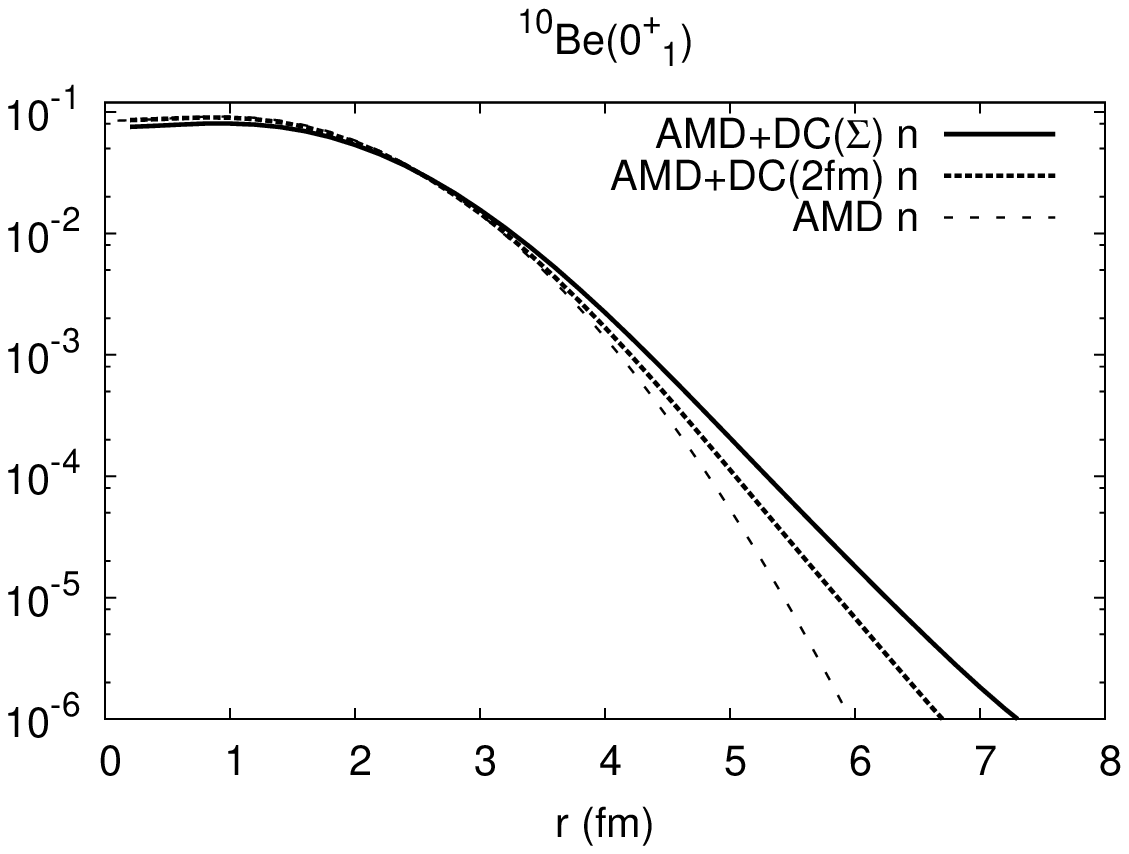}} \\
\begin{tabular}{cc}
\hspace{-5mm}
{\includegraphics[scale=0.55]{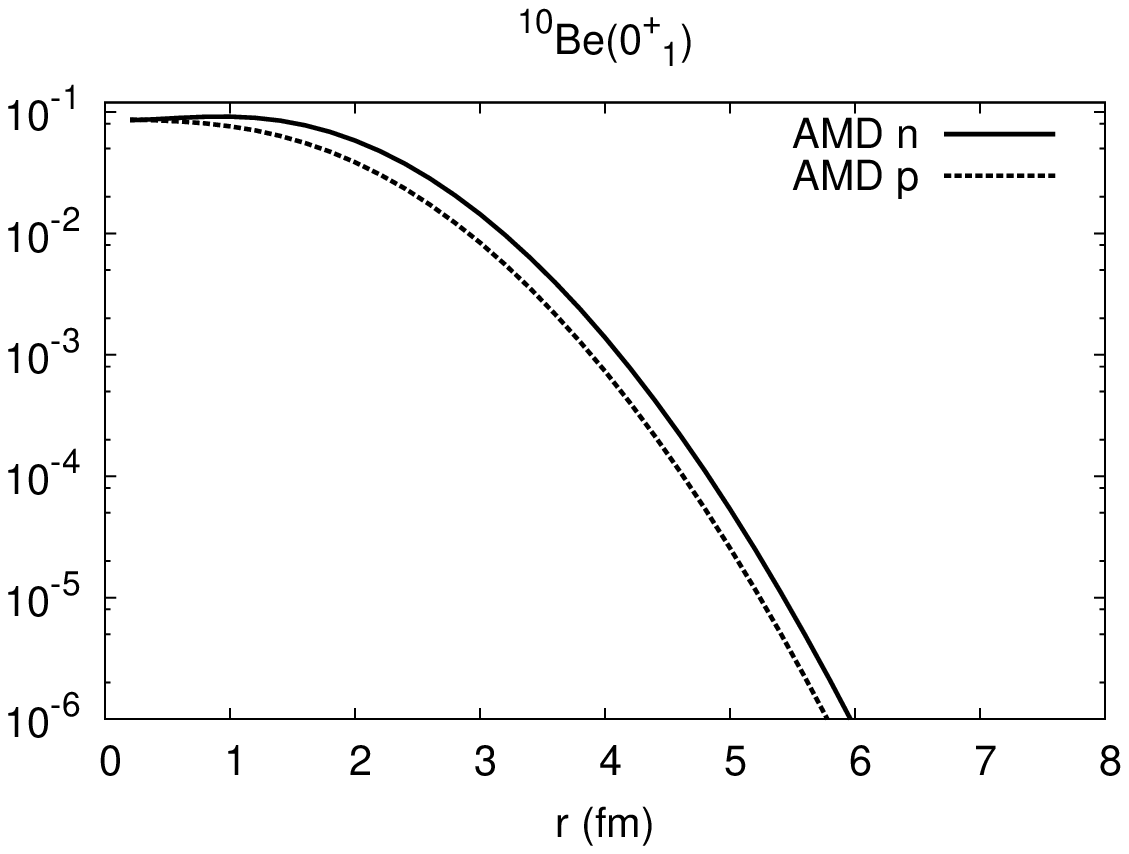}} &
\hspace{-5mm}
{\includegraphics[scale=0.55]{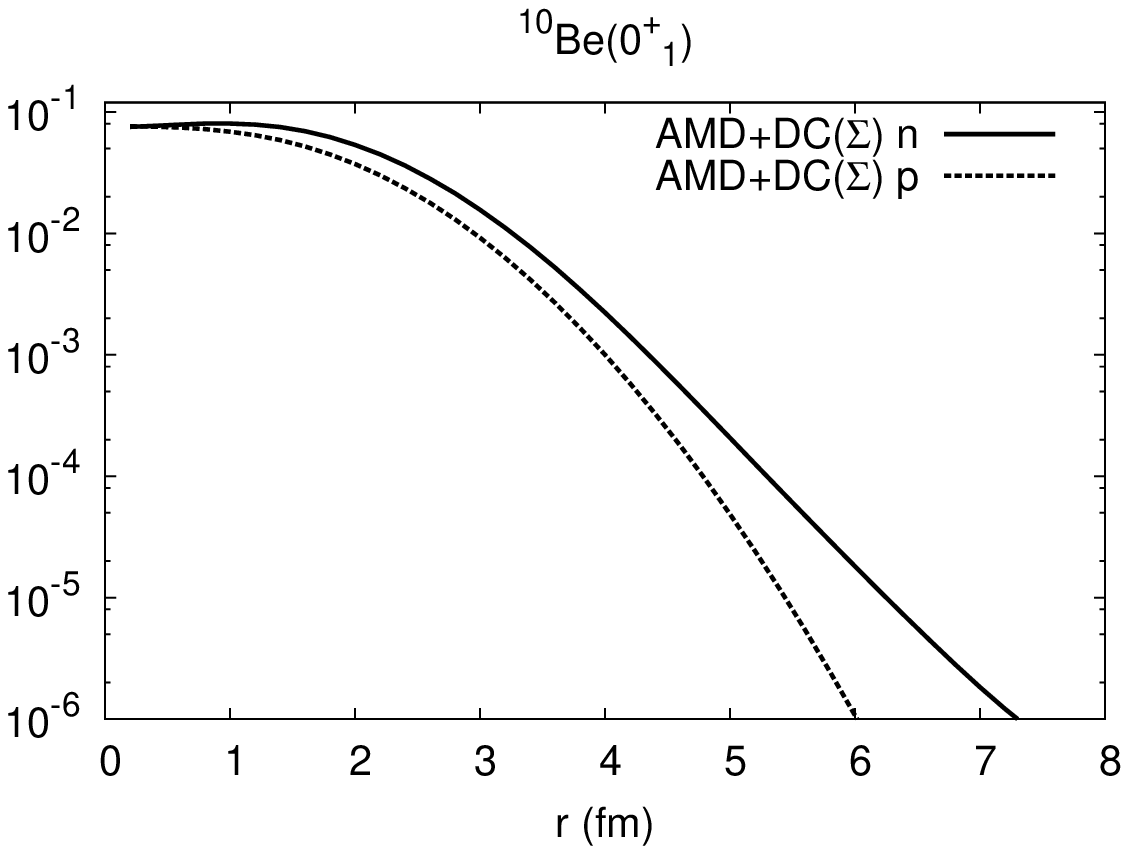}} \\
\end{tabular}
\caption{\footnotesize
Top; the neutron density calculated with AMD, AMD+DC($d=2$), AMD+DC($\Sigma_d$).
Bottom-left; the density of proton and neutron calculated with AMD.
Bottom-right; those calculated with AMD+DC($\Sigma_d$).}
\label{fig:Be10_0+1_one-particle_density}
\end{center}
\end{figure}

\begin{figure}[htbp]
\begin{center}
{\includegraphics[scale=0.55]{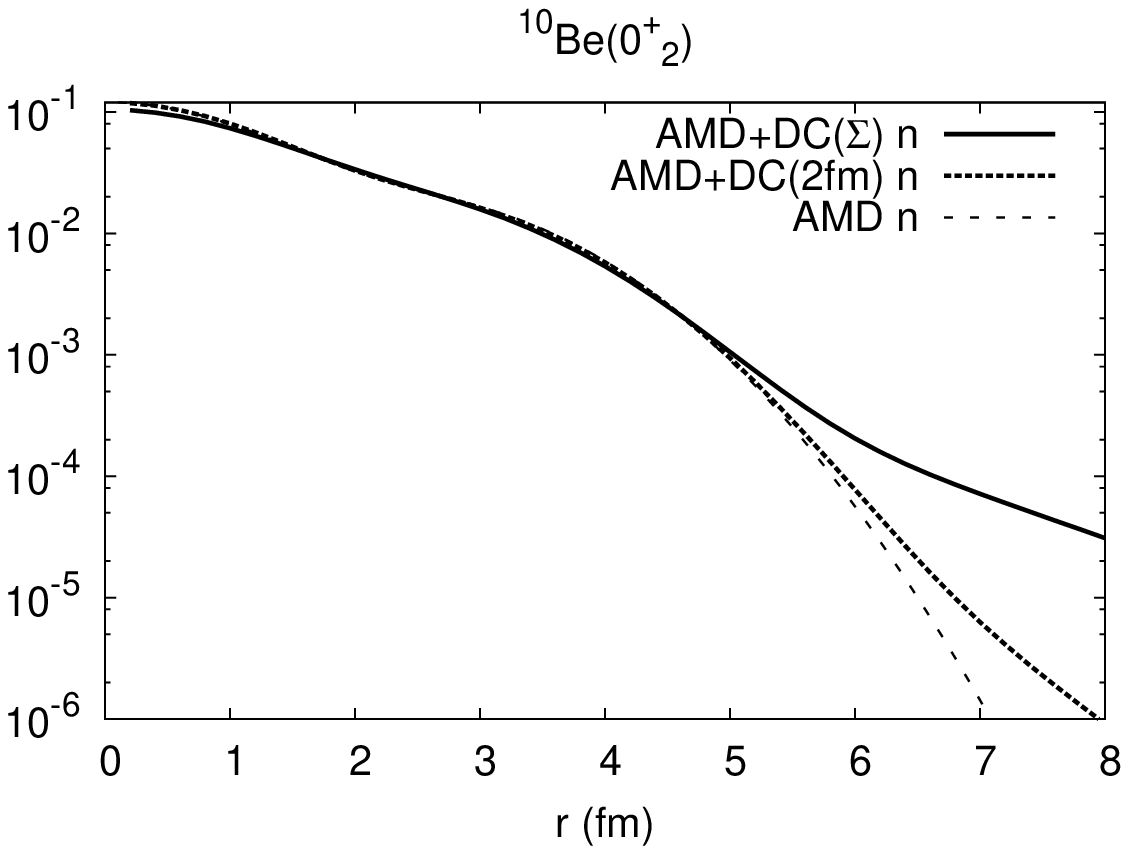}} \\
\begin{tabular}{cc}
\hspace{-5mm}
{\includegraphics[scale=0.55]{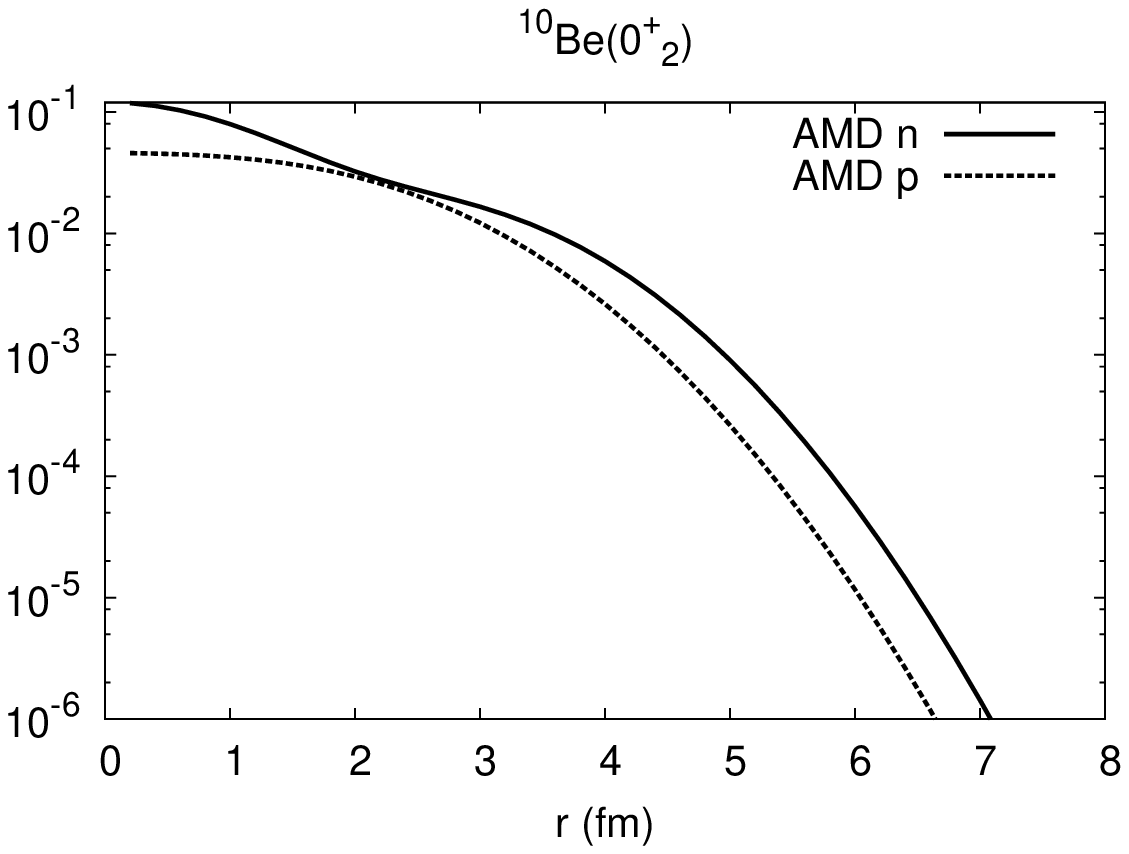}} &
\hspace{-5mm}
{\includegraphics[scale=0.55]{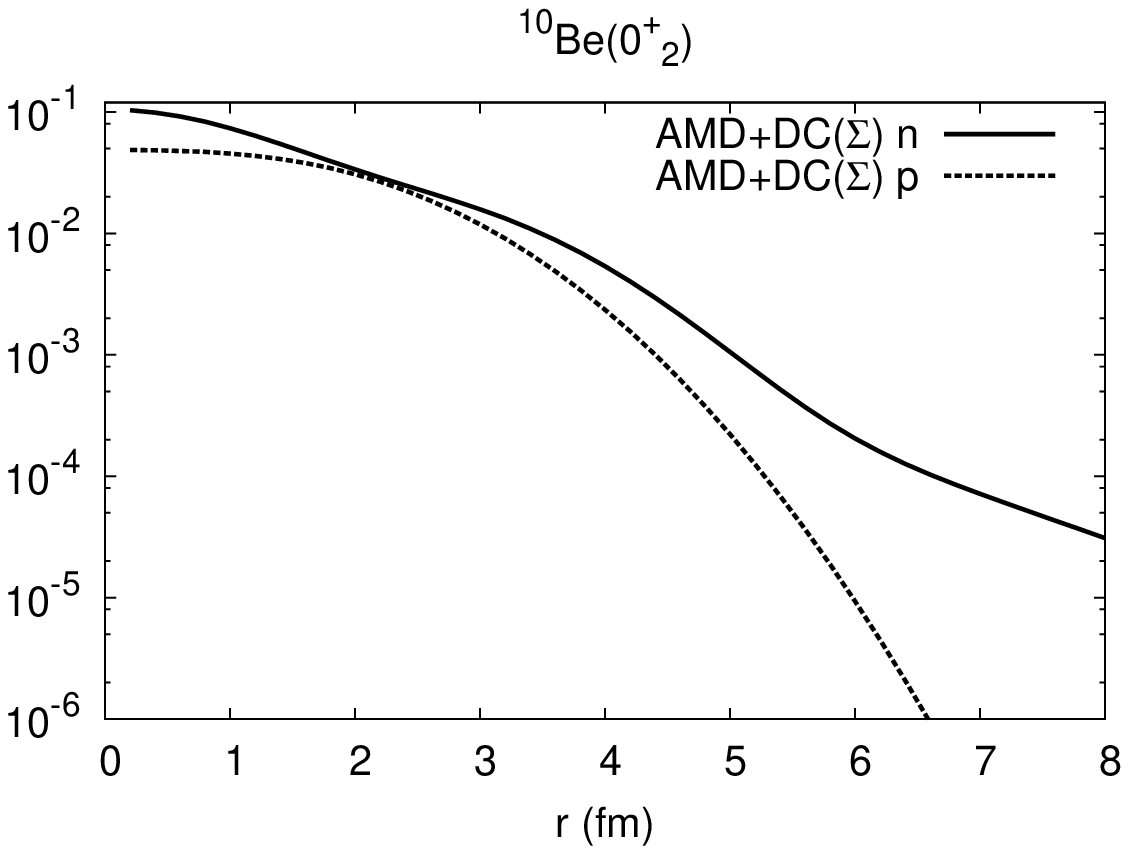}} \\
\end{tabular}
\caption{\footnotesize
Top; the one-neutron density in the $0^+_2$ state
calculated with AMD, AMD+DC($d=2$), AMD+DC($\Sigma_d$).
Bottom-left; the density of proton and neutron in the $0^+_2$ state calculated with AMD.
Bottom-right; those calculated with AMD+DC($\Sigma_d$).}
\label{fig:Be10_0+2_one-particle_density}
\end{center}
\end{figure}

We calculate the root mean square (r.m.s.) radii of the $0^+_1$ and $0^+_2$ states
to discuss the expansion of the neutron ditribution. 
The r.m.s. matter radius $r_{\rm rms}$ 
for a normalized wave function $|\Psi \rangle$ is defined as 
\begin{equation}
(r_{\rm rms})^2 = \langle r^2 \rangle = \frac{1}{A} \sum_{i=1}^A
\langle \Psi| r_i^2 |\Psi \rangle.
\end{equation}
The ones of proton and neutron are defined in the same manner. 
We also calculate the proton and neutron density.
The proton (neutron) density, $\rho_{p(n)}$, 
defined as a function of the distance from the core is defined as follows. 
\begin{equation}
\rho_{p(n)}(r) = \langle \Psi| \sum_{i=1}^{Z(N)} 
\delta(r - r_i) \delta_{\tau_i, p(n)} |\Psi \rangle.
\end{equation}
The calculated r.m.s. radii are listed in Table.~\ref{tab:Be10_r_rms}, 
and the density is shown in Figs.~\ref{fig:Be10_0+1_one-particle_density} and
\ref{fig:Be10_0+2_one-particle_density}.

In the $0^+_1$ state, the radii of AMD+DC($\Sigma_d$) increase relative to those of AMD. 
Since not only the radius of neutron but also that of proton does become large, 
the radii increases mainly due to the $\alpha$ cluster development.
Comparing the neutron density of AMD, AMD+DC($d=2$) and AMD+DC($\Sigma_d$), 
shown in Fig.~\ref{fig:Be10_0+1_one-particle_density}, 
it is seen that the neutron density 
in the inner region of the nucleus ($r \lesssim 2$ fm) is not so different 
between these wave functions. 
As seen in the $r > 4$ region, 
the expansion of the neutron density can be described by superposing the DC wave functions.
However, the difference between the proton and neutron density is not so remarkable
as the neutron-halo or skin structure.

Subsequently, we discuss the radii and the density of the $0^+_2$ state.
It can be seen that the $0^+_2$ state has a rather extensive structure
compared with the $0^+_1$ state. 
It is interesting that the radius of neutron increases remarkably 
though that of proton decreases slightly. 
We can see such an expansion of a neutron, also in the neutron density
of $^{10}$Be($0^+_2$) shown in Fig.~\ref{fig:Be10_0+2_one-particle_density}. 
The expansive structure of the valence neutrons described 
by the mixture of DC wave functions 
is well represented also in comparison of the proton and neutron density 
(the bottom-right panel of Fig.~\ref{fig:Be10_0+2_one-particle_density}). 
The neutrons are distributed rather extensively 
compared with protons, different from the $0^+_1$ state. 
However, as we discussed later, this extensive neutron structure is not necessarily 
due to the components of strongly correlated two neutrons distributed to the outer region 
but due to mainly those of uncorrelated two neutrons.

\begin{figure}[htbp]
\begin{center}
\begin{tabular}{cc}
\hspace{-7mm}
{\includegraphics[scale=0.6]{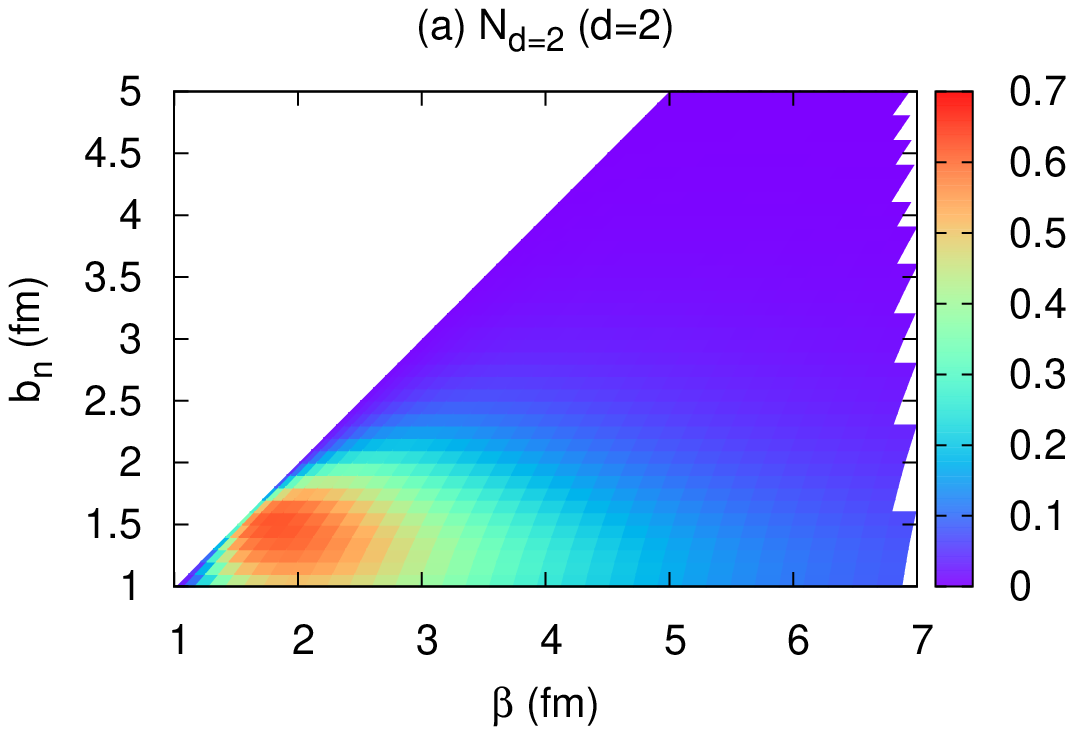}} &
\hspace{-10mm}
{\includegraphics[scale=0.6]{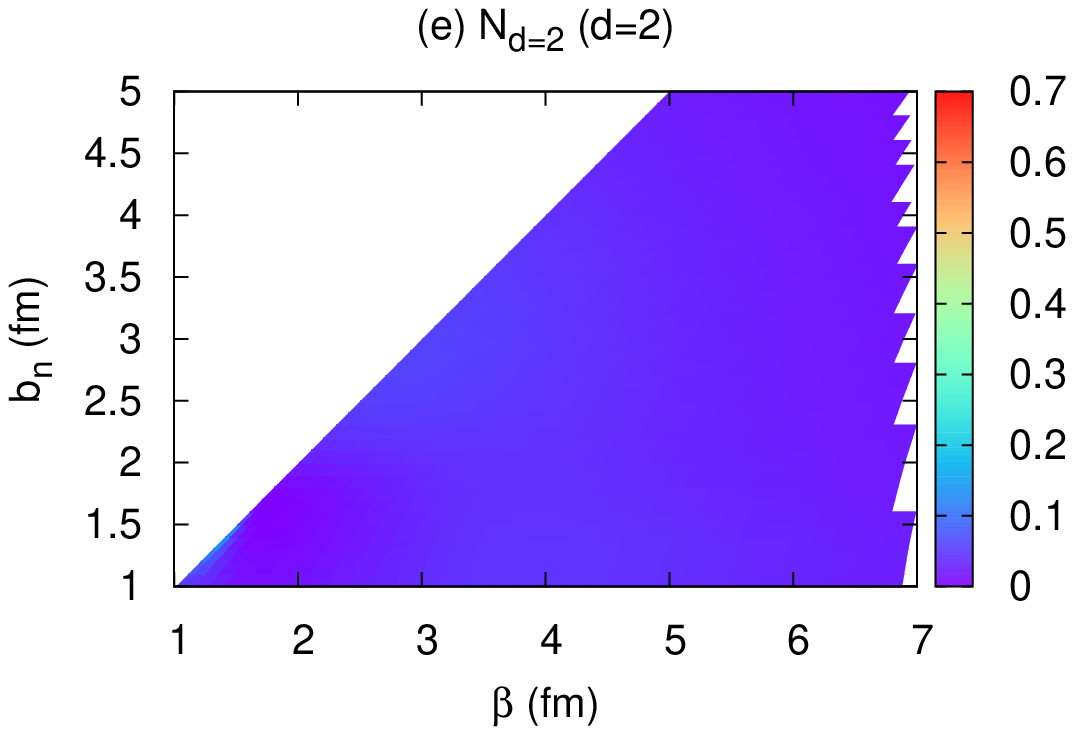}}
\vspace{-6mm} \\
\hspace{-7mm}
{\includegraphics[scale=0.6]{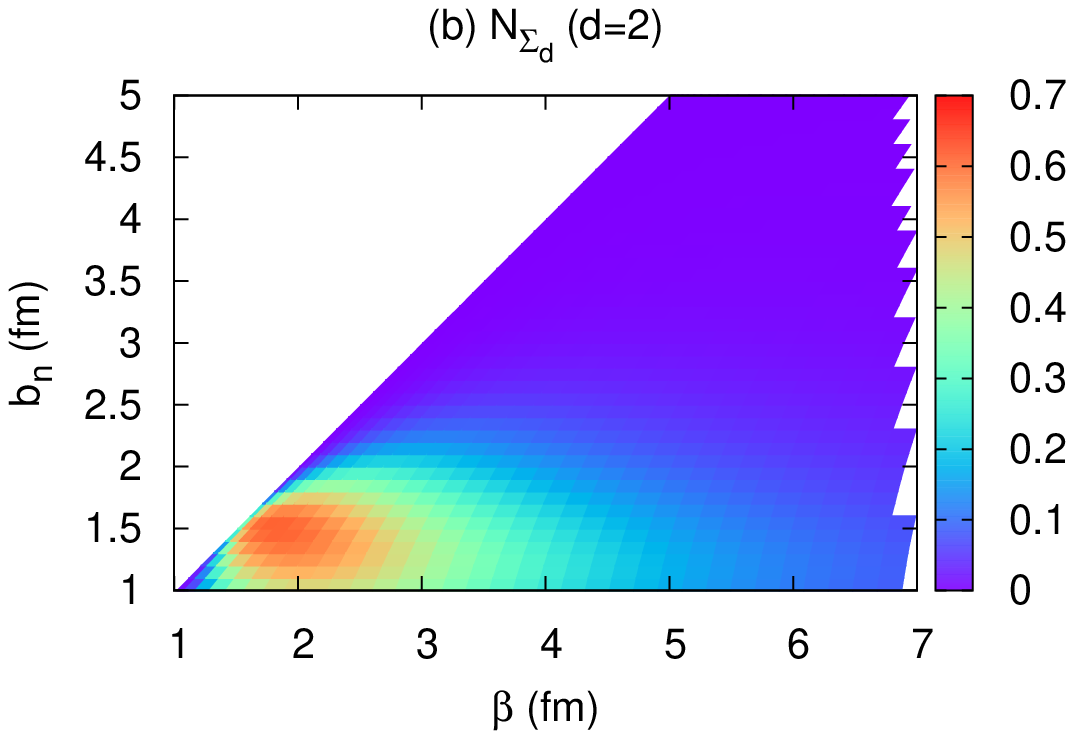}} &
\hspace{-10mm}
{\includegraphics[scale=0.6]{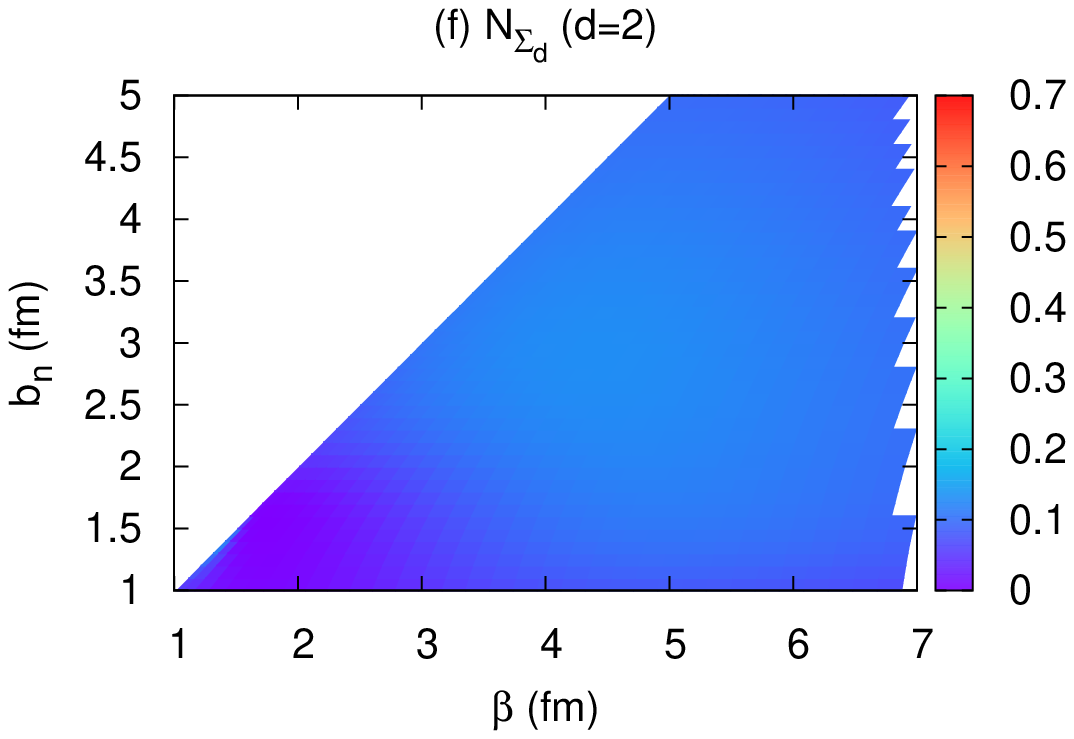}}
\vspace{-6mm} \\
\hspace{-7mm}
{\includegraphics[scale=0.6]{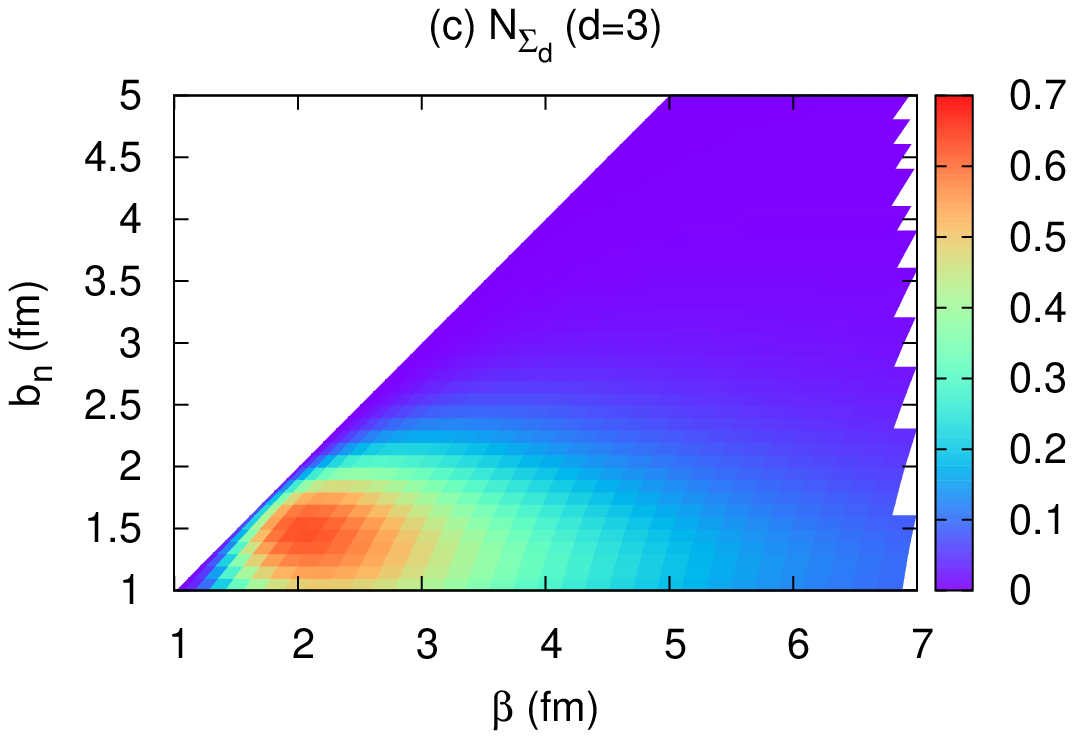}} &
\hspace{-10mm}
{\includegraphics[scale=0.6]{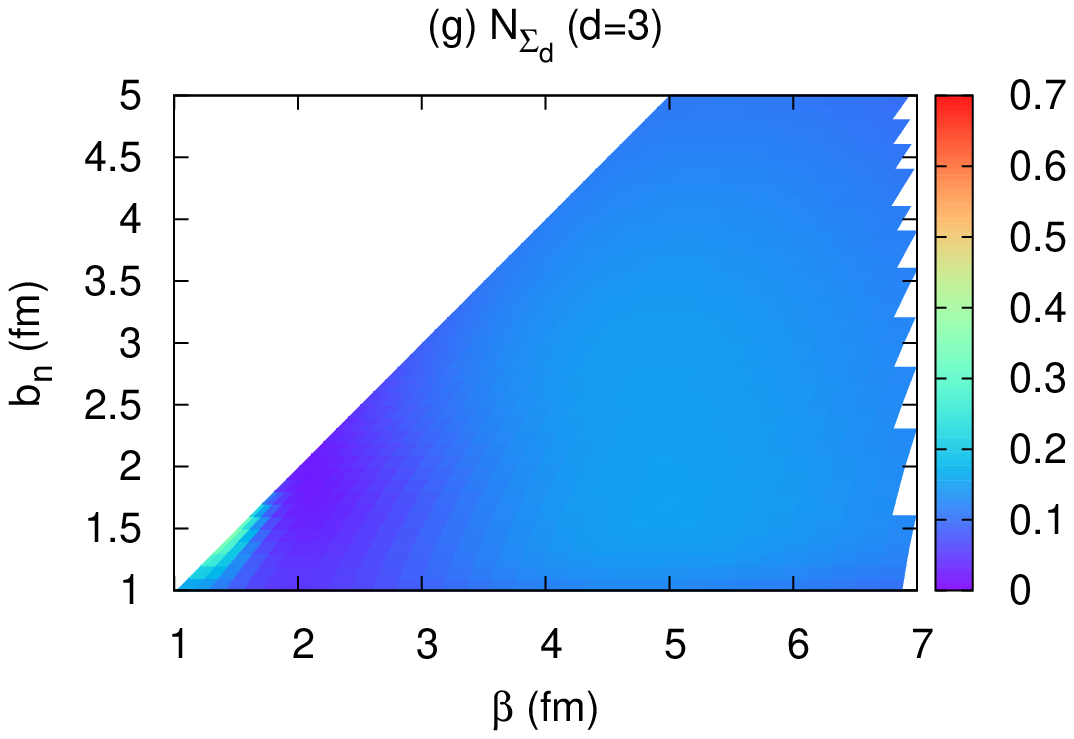}} 
\vspace{-6mm} \\
\hspace{-7mm}
{\includegraphics[scale=0.6]{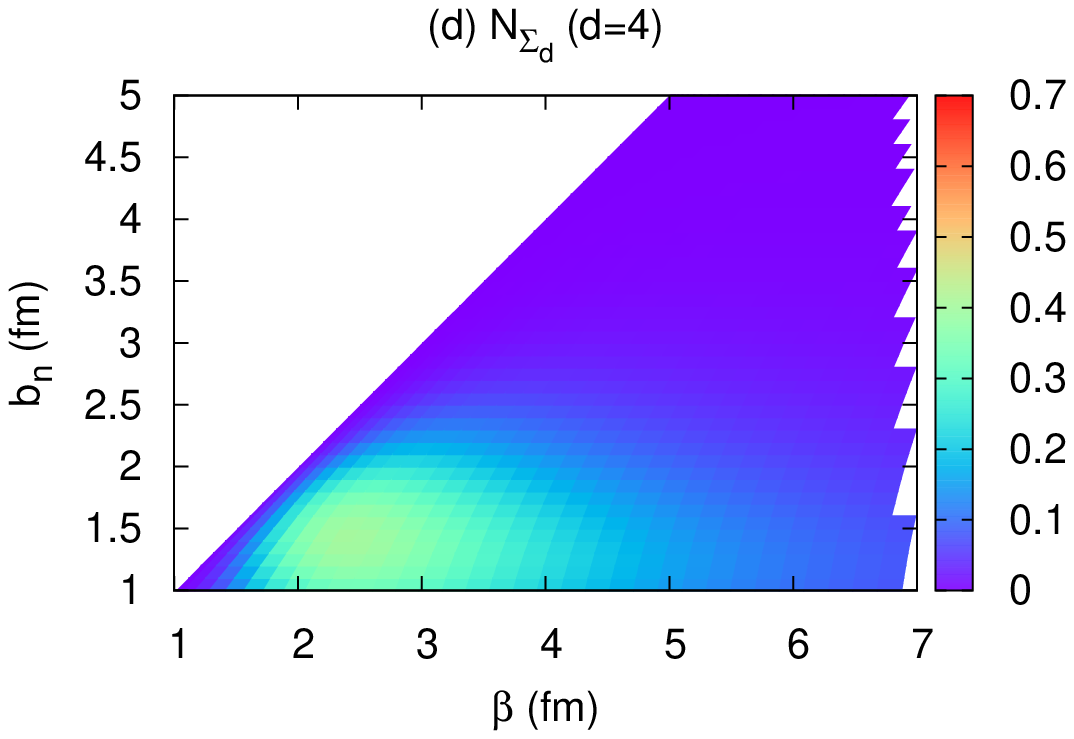}} &
\hspace{-10mm}
{\includegraphics[scale=0.6]{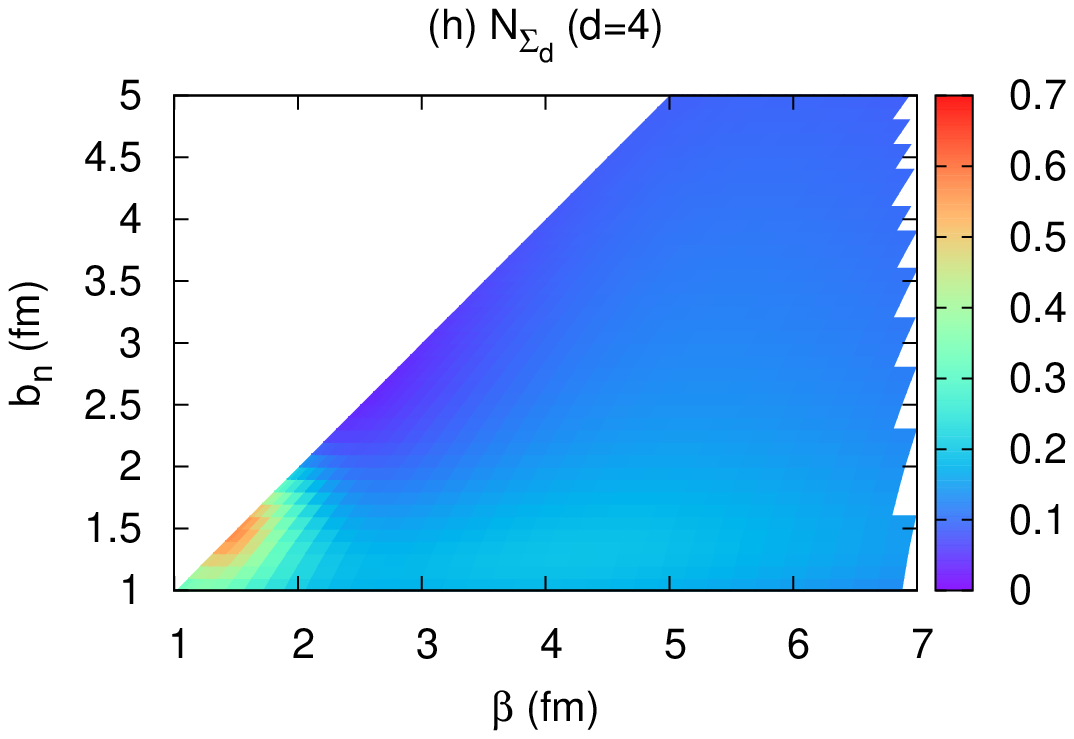}} \\
\end{tabular}
\vspace{-3mm}
\caption{\footnotesize
(a)-(d); the overlap of the $0^+_1$ state with DC wave function 
whose parameter $d$ is fixed
(the fixed value is represented in the parenthesis) on the $\beta$-$b_n$ surface. 
The label $N_{d=2}$, or $N_{\Sigma_d}$, indicates the overlap of the $0^+_1$ state
obtained by the AMD+DC($d=2$), or the AMD+DC($\Sigma_d$) wave function. 
They are explained in the text.
(e)-(h); the overlap of the $0^+_2$ in the same manner as the $0^+_1$. }
\label{fig:Be10_overlap_0+}
\end{center}
\end{figure}

\subsection{Dineutron component in $^{10}$Be($0^+_1$) and $^{10}$Be($0^+_2$)}
In the AMD+DC calculations, the $0^+_1$ and $0^+_2$ states 
contain the dominant AMD wave functions with the mixture of DC wave functions.
In the $0^+_1$ state of AMD+DC($d=2$), the overlap with AMD is $97$\%, 
and the contribution of DC wave functions is minor. 
On the other hand, the AMD component in the $0^+_1$ state of AMD+DC($\Sigma_d$) 
decreases to $89$\%. 
In the $0^+_2$ state, the AMD component in AMD+DC($d=2$) and 
AMD+DC($\Sigma_d$) is $98$\% and $82$\%, respectively.
Therefore, the mixture of the dineutron component described with the DC wave functions 
is enhanced by superposing various $d$-values, i.e. considering the 2$\alpha$ relative motion. 
However the AMD component is still dominant and 
we suppose that the main properties of the AMD+DC states may not 
be so much different from those of the AMD states. 
Despite its a little mixing amplitude, 
the mixture of the DC component can significantly contribute to
the energy of the system and the dineutron component far from the core. 
In the following, we analyze the dineutron component in these states in detail.

To investigate the dineutron component, 
we calculate the overlap of the AMD+ DC($d=2$) or AMD+DC($\Sigma_d$) wave function
with the DC wave function whose parameters $b_n, \beta$ and $d$ are fixed to certain values. 
The overlap of a normalized wave function $|\Psi \rangle$ 
with a DC wave function $|\Phi_{\rm DC} (d, \beta, b_n) \rangle$ is written as 
\begin{equation}
N(d,\beta,b_n) = | \langle \Psi|\Phi_{\rm DC} (d, \beta, b_n) \rangle |^2. 
\end{equation}
This quantity is an indicator 
which shows how the dineutron component contributes to the state.
We denote the overlap of AMD+DC($d=2$) 
and that of AMD+DC($\Sigma_d$) with $\Phi_{\rm DC}(d,\beta,b_n)$
as $N_{d=2}(d,\beta,b_n)$ and $N_{\Sigma_d}(d,\beta,b_n)$, respectively.
The calculated values of $N_{d=2}(d,\beta,b_n)$ for $d=2$ fm 
and those of $N_{\Sigma_d}(d,\beta,b_n)$ for $d=2,3,4$ fm 
are plotted on the $\beta$-$b_n$ plane in Fig.~\ref{fig:Be10_overlap_0+}.
The panels in the left row of Fig.~\ref{fig:Be10_overlap_0+} 
show the overlaps of the $0^+_1$ state, 
and those in the right row show the overlaps of the $0^+_2$ state. 
Since we deal with the parameter $b_n$ as the size of the dineutron 
and $\beta$ as the spread of the dineutron distribution from the core, 
we consider the amplitude in the small $b_n$ and relatively large $\beta$ region as 
the component of a developed compact dineutron. 

At first, we discuss the dineutron component in the $0^+_1$ state. 
As shown in Fig.~\ref{fig:Be10_overlap_0+}(b)-(d), 
AMD+DC($\Sigma_d$) contains a large component of the DC wave function whose $d$ is 2, 3 fm. 
The rather large component near $(\beta, b_n) \sim (2.0, 1.5)$ 
(the pocket of the dineutron energy, see Fig.~\ref{fig:E_B_b})
corresponds to the rather compact dineutron at the nuclear surface.
Seeing these overlaps, 
there is the tail characterized by the relatively gradual decrease 
from the peak $(\beta, b_n) \sim (2.0, 1.5)$
toward the large $\beta$ region along $b_n \sim 1.5$ fm.
This component is the dineutron distributed 
away from the core to some extent keeping a compact size.
Interestingly, the significant amplitude of the compact dineutron with $b_n \sim 1.5$ fm
in the large $\beta$ region can be seen even in $N_{\Sigma_d}(d=4)$ 
for the spatially developed two $\alpha$ clusters.
However, the maximum overlap with the DC wave function having $d=4$ fm 
is half of those for $d=2,3$ fm, 
which means that the component of the extremely developed two $\alpha$ clusters is not dominant 
in the $0^+_1$ state, unlike the $0^+_2$.

\begin{figure}[b]
\begin{center}
\begin{tabular}{cc}
\hspace{-5mm}
{\includegraphics[scale=0.55]{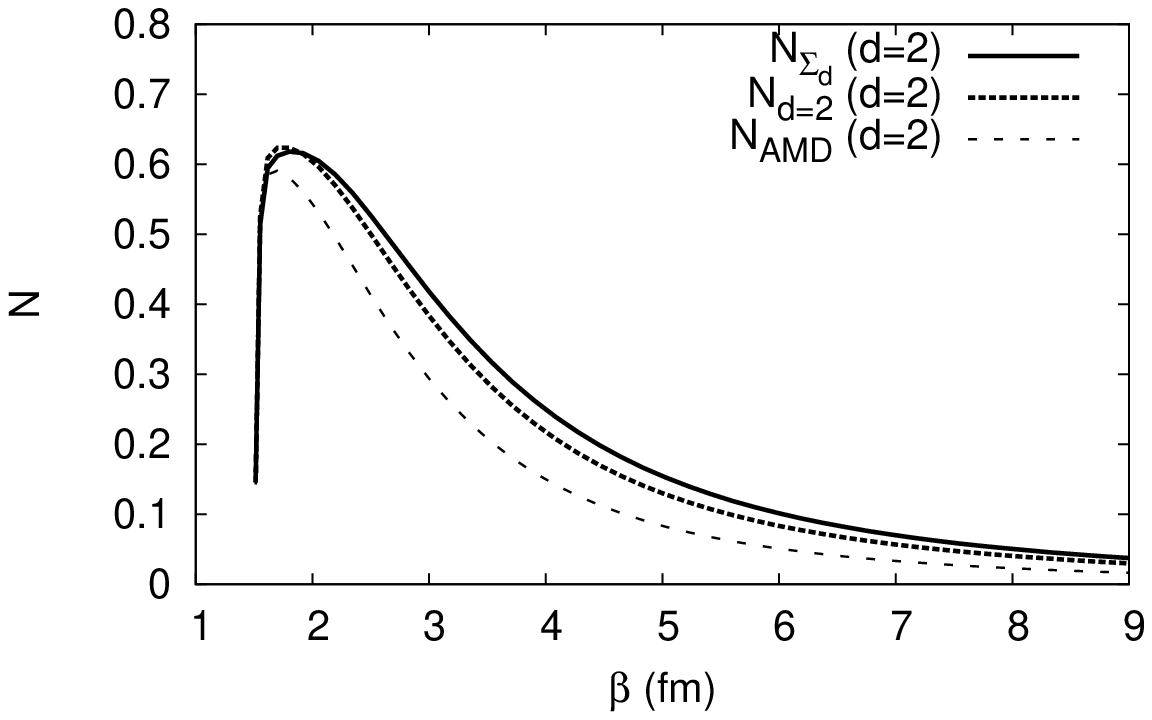}} &
\hspace{-5mm}
{\includegraphics[scale=0.55]{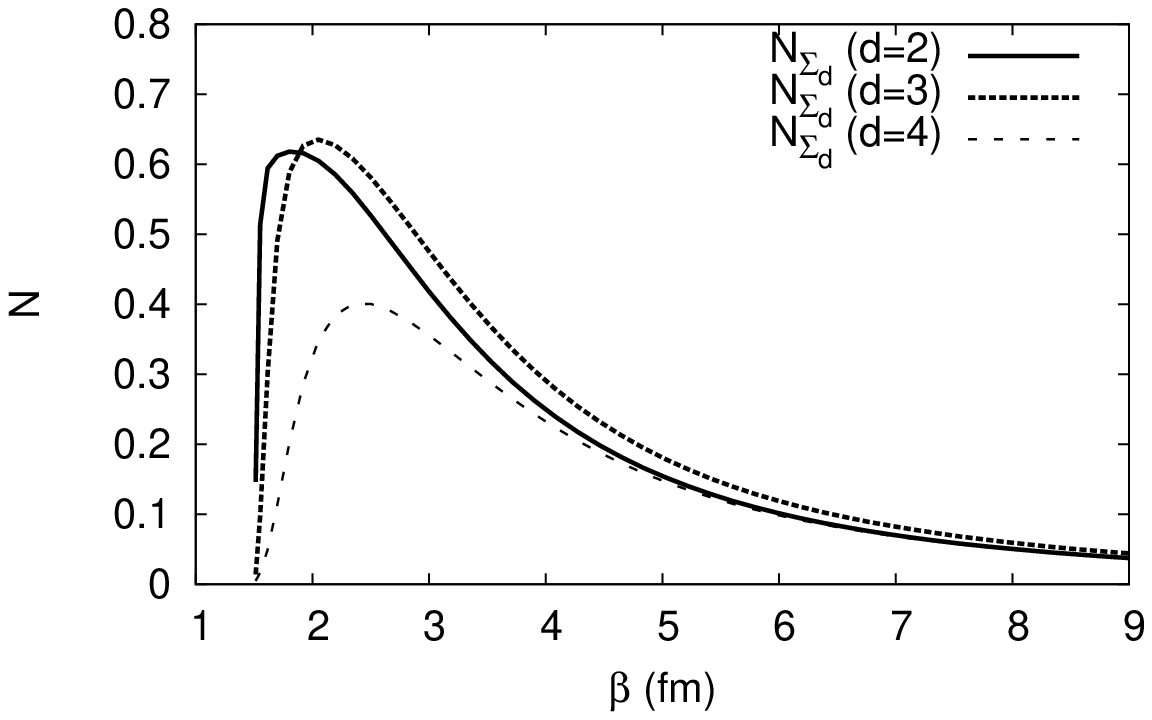}} \\
\end{tabular}
\vspace{-5mm}
\caption{\footnotesize
Left; the overlap of the $0^+_1$ states with the DC wave function 
where $b_n=1.5$ fm and $d=2$ fm.
The label $N_{\rm AMD}$, $N_{d=2}$, or $N_{\Sigma_d}$ indicates 
the overlap of the $0^+_1$ states obtained by 
the AMD, AMD+DC($d=2$), or AMD+DC($\Sigma_d$) wave function.
Right; the overlap of the $0^+_1$ states 
described with the AMD+DC($\Sigma_d$) wave function 
with the DC wave function having $b_n$ fixed to 1.5 fm and $d$ to 2, 3, 4 fm.}
\label{fig:Be10_overlap_b1.5}
\end{center}
\end{figure}

To see how far the compact dineutron is distributed from the core, 
we show the overlap $N(d,\beta,b_n)$ with the fixed dineutron size $b_n=1.5$ fm 
in Fig.~\ref{fig:Be10_overlap_b1.5}.
In the left figure of Fig.~\ref{fig:Be10_overlap_b1.5}, 
the overlaps, $N_{d=2}$ and $N_{\Sigma_d}$, of AMD+DC($d=2$) and AMD+DC($\Sigma_d$)
are enhanced compared to that of AMD 
in the region $\beta > 2$ fm, where the dineutron is distributed away from the core. 
The remarkable increase of the overlap in this region indicates that 
the mixture of the DC wave functions plays a significant role in describing the developed dineutron. 
Next, we compare $N_{d=2}$ and $N_{\Sigma_d}$ to see the effect of core excitation. 
The maximum of each overlap exists at $\beta \sim 2$ fm, 
which corresponds to the vicinity of the nuclear surface, 
and there is no pronounced difference between them near this point.
That means the variability of the $\alpha$-$\alpha$ distance 
does not affect the development of the dineutron correlation 
from the inside to the surface of $^{10}$Be. 
The difference is seen in the region $\beta > 2$ fm, 
where the dineutron is distributed distantly from the core to some extent. 
$N_{\Sigma_d}$ is a bit larger than $N_{d=2}$ there, 
so that this increase suggests that 
the core excitation, such as the relative motion of clusters, 
may contribute significantly to the enhancement of the dineutron correlation. 

We show the overlap $N_{\Sigma_d}(d,\beta,b_n)$ with the fixed dineutron size $b_n=1.5$ fm 
for DC wave functions having various $d$-values in Fig.~\ref{fig:Be10_overlap_b1.5}. 
The dineutron component near the nuclear surface ($\beta \sim 2$ fm) is 
greater in the case $d=2, 3$ fm than $d=4$ fm. 
The value of $\beta$ giving the maximum of $N_{\Sigma_d}$ becomes larger 
as the parameter $d$ of the overlapped DC wave function goes from 2 fm to 4 fm. 
That means the dineutron develops spatially and becomes distributed away from the core
in accordance with the development of two $\alpha$ clusters. 

Subsequently, we consider the dineutron component in $^{10}$Be($0^+_2$).
As seen in Fig.~\ref{fig:Be10_overlap_0+}(f)-(h), 
the $0^+_2$ state has the largest overlap with the DC wave function having $d=4$ fm 
at $(\beta,b_n) \sim (1.5,1.5)$, 
which corresponds to not the compact dineutron near the nuclear surface 
but two neutrons distributed in the middle of the distant two $\alpha$ clusters.
This behavior is consistent with the molecular orbital structure 
with the $\sigma^2$ configuration, 
which has the nodal structure along the axis 2$\alpha$ located on.
In any cases of $d=2$, 3 and 4 fm, 
the overlaps have not too small amplitudes in the almost whole region, 
where the distance between two neutrons and their spread from the core are large.  
Namely, in the present calculation, two valence neutrons in the $0^+_2$ state
are distributed extensively from the core with less correlation.  

\subsection{Candidate for the resonance state $0^+_3$}
\begin{figure}
\begin{center}
\begin{tabular}{cc}
{\includegraphics[scale=0.55]{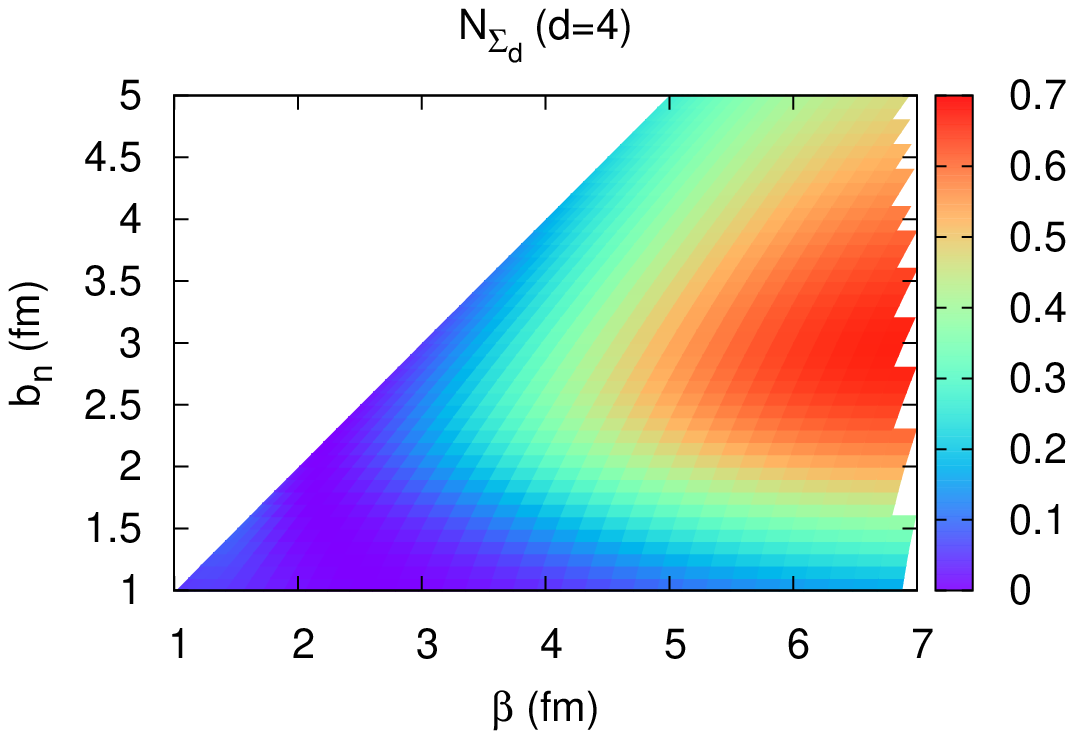}}&
\hspace{-7mm}
{\includegraphics[scale=0.55]{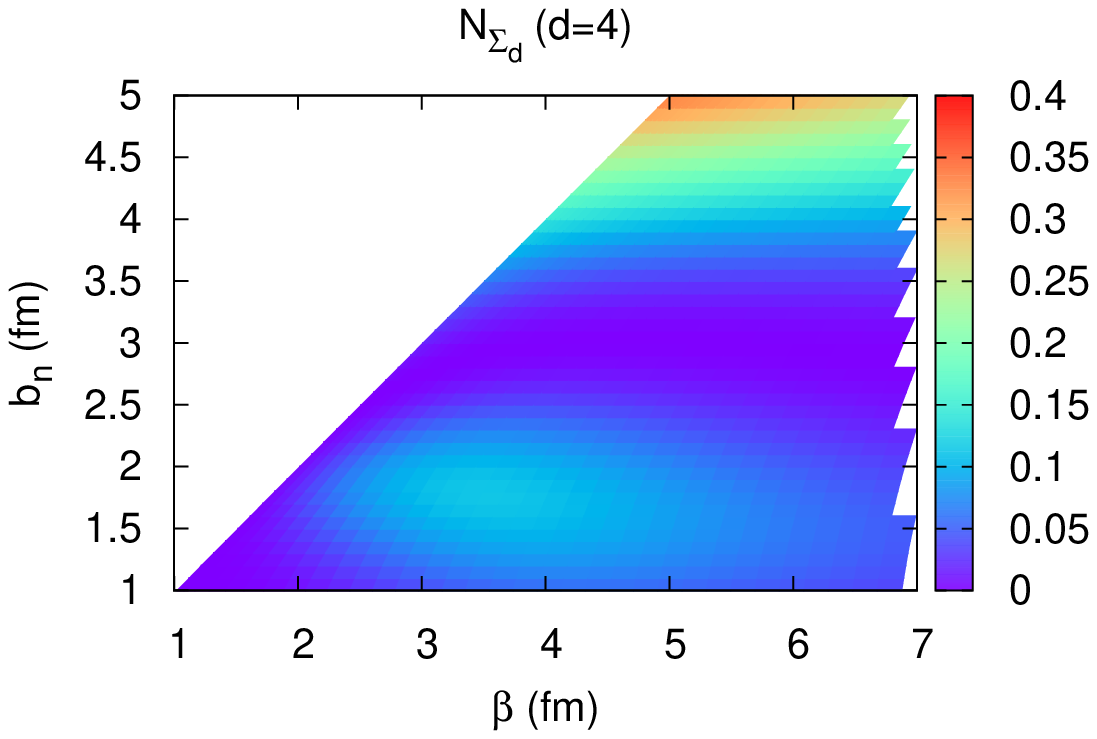}} \vspace{-3mm} \\ 
{\includegraphics[scale=0.55]{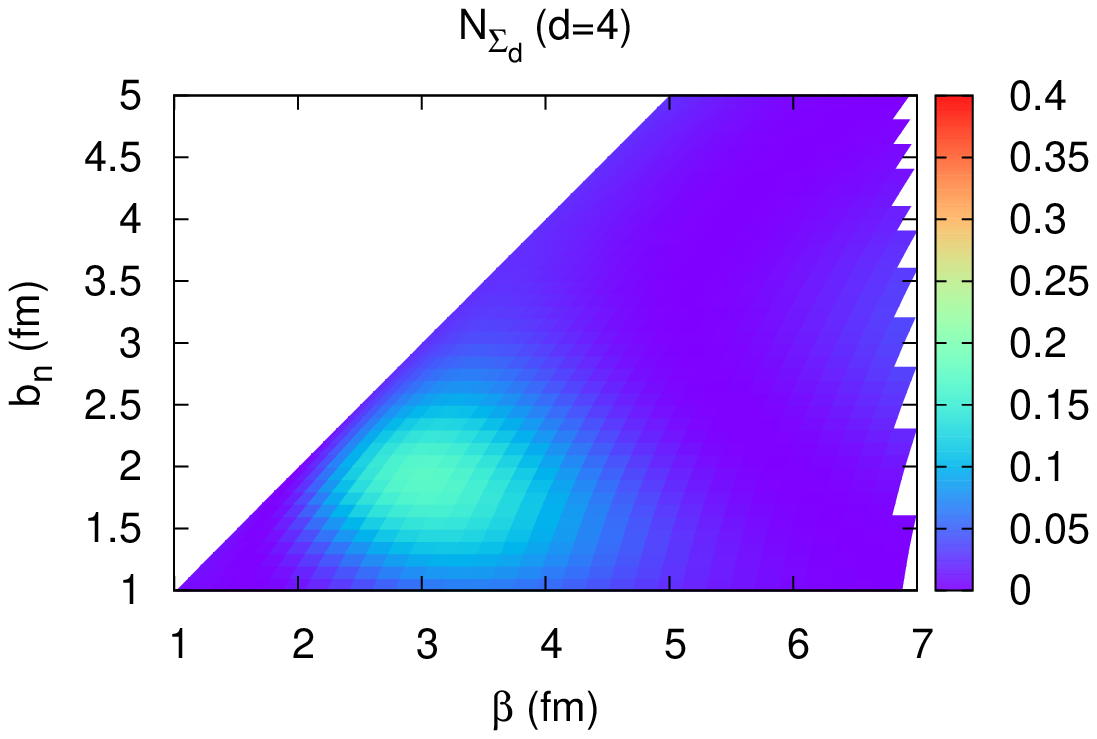}} &
\end{tabular}
\vspace{-5mm}
\caption{\footnotesize 
Top-left; the overlap of the third lowest state in the AMD+DC($\Sigma_d$) 
with the DC wave function having $d=4$.
Top-right; that of the forth lowest state.
Bottom; that of the fifth lowest state,
which is expected to be a resonance state.
Note that the scales of amplitude in the figure of the forth and fifth states 
are different from so far.}
\label{fig:Be10_overlap_0+3}
\end{center}
\end{figure}

In the preceding studies, the $0^+_3$ state having a developed dineutron
around two developed $\alpha$ clusters were predicted \cite{itagaki00,suhara10}, 
though there is no experimental evidence yet. 
In the present calculation, many states besides the $0^+_1$ and $0^+_2$ states are obtained 
in the AMD+DC wave functions. 
Although most of them are the insignificant continuum states, 
there would be a resonance-like state with a developed dineutron component. 
In Fig.~\ref{fig:Be10_overlap_0+3}, 
we show the overlap, $N_{\Sigma_d}$, of the AMD+DC($\Sigma_d$) 
with the DC wave function for typical examples corresponding to continuum states, 
and that for the candidate of a resonance state. 
We choose the parameter $d=4$ for the overlapped DC wave function, 
for the resonance state is thought to contain the clusters bound loosely each other  
so that the overlap should be large when two $\alpha$s are distant.  

The overlap of the third lowest state in AMD+DC($\Sigma_d$) 
shown in the top-left panel of Fig.~\ref{fig:Be10_overlap_0+3}
is the typical one for a continuum state.
This state has an enormous overlap 
with the DC whose both parameters $b_n$ and $\beta$ are large. 
This tendency reflects that the three-body unbound feature of two valence neutrons and the core. 
Another example is shown in the top-right of Fig.~\ref{fig:Be10_overlap_0+3}
for the forth lowest one in AMD+DC($\Sigma_d$). 
Two peaks at $b_n \sim 1.7$ and $5$ fm 
and a node at $b_n \sim 3$ fm are seen in this figure.   
This also shows a feature of a continuum state. 

The state shown in the bottom panel of Fig.~\ref{fig:Be10_overlap_0+3} is 
the overlap of the fifth lowest state, 
which we expect to be the candidate for a resonance state. 
This state have a peak in the moderate $\beta$ and $b_n$ region $(\beta, b_n) \sim (3,2)$, 
which represents a relatively smaller distance between two neutrons
and that between the dineutron and the core than 
those in the cases of continuum states mentioned above. 
The excited energy of this state is close to the ones 
calculated with different frameworks in preceding studies 
\cite{itagaki00,suhara10} (see Fig.~\ref{fig:Be10_excited_energy_spectrum}). 
We, however, cannot yet conclude that this fifth state is a resonance 
with only the evidence we have shown here. 
Since this state may be combined with contiuum states, 
the matter radius and density do not converge. 
So we need to analyze it more closely in a different manner in future.

\section{Summary}

We have proposed a new approach to investigate the dineutron correlation. 
To incorporate the dineutron component, 
we introduce ``the dineutron condensate wave function'', for short, the DC wave function. 
The wave function can describe a system composed of 
a deformative core and surrounding dineutrons which condensate in the lowest $S$-orbit.
In the AMD+DC method for the investigation of the dineutron correlation in neutron-rich nuclei, 
DC wave functions are combined with AMD wave functions.

We have applied the DC wave function to a particular system 
which consists of a 2$\alpha$ core and one dineutron. 
Analyzing the dineutron size dependence of the energy of the 2$\alpha$+2n system, 
we discuss the mechanism of the dineutron formation around the 2$\alpha$ core.
We have found that the dineutron favors a relatively compact size
when distributed near the nuclear surface 
because of the Pauli blocking effect from the core. 
Core effects are essential for the dineutron formation, 
and this mechanism may lead the dineutron correlation enhancement in general nuclei. 

We have investigated the dineutron correlation in $0^+$ states of $^{10}$Be 
with the AMD+DC method.
The obtained wave functions for the $0^+_1$ and $0^+_2$ states 
have the dominant AMD component with slightly mixing of the DC component.
The mixing component of the DC wave functions enhances 
when we take into account the core excitation by superposing DC wave functions 
with different $\alpha$-$\alpha$ distances.
In the $0^+_1$ state, we have found a tail structure of the dineutron 
in the region far from the core.
In spite of the dineutron tail, 
the component of the strongly correlated two neutrons
does not contribute so much to such one-body observables 
as the neutron radius and density, unlike halo nuclei.
It may be because two neutrons are somehow strongly bound in the $^{10}$Be system.

In addition to analyzing the dineutron component in the $0^+_1$ and $0^+_2$ states, 
we have searched for the $0^+_3$ state having a developed dineutron, 
which were predicted theoretically by  preceding studies \cite{suhara10,itagaki00}. 
We have found a candidate for the state in the present results. 
We, however, could not conclude that the candidate state is a resonance 
because it contains the component of continuum states 
and the radius and density do not converge.
We need further investigations by dealing with resonant and continuum states properly.

In the present work, we have applied the new framework,
``the DC wave function'', to a virtual system 2$\alpha$+2n, 
and its practical application, ``the AMD+DC method'', 
to a simple system $^{10}$Be. 
We still have a room for the improvement of this framework, 
for example adopting a deformed distribution of dineutrons. 
However, even in the simplest application of this work, 
we have obtained important conclusion for the mechanism of the dineutron formation
and the possibility of the effect of the core structure 
to enhance the dineutron correlation. 
And also we have confirmed the utility of this framework 
to describe the dineutron correlation in the nucleus 
which contains a deformed and excited core. 
We expect that our framework can be applied to various nuclei 
without critical assumptions. 

In future, we will apply the AMD+DC method to various nuclei, 
in particular neutron-rich nuclei which are characterized with 
exotic structures such as the neutron halo and
are expected to contain a developed dineutron. 
For instance, $^{14}$Be is a challenging problem 
since it may be a two-neutron-halo nucleus with a deformed $^{12}$Be core. 
One of our final objectives is to 
clarify the universal properties of the dineutron correlation
by systematic investigations of the dineutron correlation in diverse circumstances. 
Besides the case that the systems include one dineutron, 
we also would like to consider the system with a few dineutrons 
and analyze the property of a dineutron condensate state
with developed dineutrons near the low-density surface.

\section*{Acknowledgments}
This work was supported by Grant-in-Aid for Scientific Research 
from Japan Society for the Promotion of Science (JSPS).
It was also supported by
the Grant-in-Aid for the Global COE Program ``The Next Generation of Physics,
Spun from Universality and Emergence'' 
from the Ministry of Education, Culture, Sports, Science and Technology (MEXT) of Japan.
A part of the computational calculations of this work was performed by using the
supercomputers at YITP and done in Supercomputer Projects
of High Energy Accelerator Research Organization (KEK).

\def\@biblabel#1{#1) }

\end{document}